\documentclass[12pt]{article}
\usepackage{amsmath}
\usepackage{amsthm}
\usepackage{amssymb}
\usepackage{amsfonts}
\usepackage{epic}
\usepackage{eepic}
 \usepackage{times}
\usepackage[matrix,arrow]{xy}
\usepackage{macros}
\usepackage{epsfig}
\usepackage{graphicx}% Include figure files
\usepackage{bm}% bold math
\theoremstyle{plain}

\theoremstyle{remark}

\newcommand{\sst}{\scriptscriptstyle}

\newcommand{\vt}{\vartheta}

\renewcommand{\1}{\one}
\renewcommand{\2}{\two}
\newcommand{\3}{\three}

\newcommand{\pa}{\partial}
\newcommand{\ot}{\otimes}
\newcommand{\frt}{{\mathfrak t}}
\newcommand{\ra}{\to}

\newcommand{\fsl}{{\mathfrak s}{\mathfrak l}}

\newcommand{\al}{\alpha}

\newcommand{\be}{\beta}
\newcommand{\ga}{\gamma}
\newcommand{\Ga}{\Gamma}
\newcommand{\de}{\delta}
\newcommand{\De}{\Delta}
\newcommand{\ep}{\epsilon}

\newcommand{\La}{\Lambda}

\newcommand{\si}{\sigma}

\newcommand{\vf}{\varphi}

\newcommand{\BBE}{E}

\newcommand{\CA}{{\mathcal A}}
\newcommand{\CB}{{\mathcal B}}

\newcommand{\CC}{{\mathcal C}}
\newcommand{\CD}{{\mathcal D}}
\newcommand{\CE}{{\mathcal E}}
\newcommand{\CF}{{\mathcal F}}
\newcommand{\CG}{{\mathcal G}}
\newcommand{\CH}{{\mathcal H}}

\newcommand{\CL}{{\mathcal L}}
\newcommand{\CM}{{\mathcal M}}
\newcommand{\CN}{{\mathcal N}}
\newcommand{\CO}{{\mathcal O}}
\newcommand{\CP}{{\mathcal P}}
\newcommand{\CQ}{{\mathcal Q}}
\newcommand{\CR}{{\mathcal R}}
\newcommand{\CS}{{\mathcal S}}
\newcommand{\CT}{{\mathcal T}}

\newcommand{\CU}{{\mathcal U}}
\newcommand{\CV}{{\mathcal V}}

\newcommand{\CX}{{\mathcal X}}
\newcommand{\CY}{{\mathcal Y}}
\newcommand{\CZ}{{\mathcal Z}}

\newcommand{\SB}{{\mathsf B}}
\newcommand{\SC}{{\mathsf C}}

\newcommand{\SF}{{\mathsf F}}

\newcommand{\SH}{{\mathsf H}}

\newcommand{\sk}{{\mathsf k}}
\newcommand{\SL}{{\mathsf L}}

\newcommand{\SM}{{\mathsf M}}

\newcommand{\SR}{{\mathsf R}}
\renewcommand{\SS}{{\mathsf S}}
\newcommand{\ST}{{\mathsf T}}
\newcommand{\SU}{{\mathsf U}}
\newcommand{\SV}{{\mathsf V}}

\newcommand{\SX}{{\mathsf X}}

\newcommand{\fe}{{\mathfrak e}}

\newcommand{\fv}{{\mathfrak v}}

\newcommand{\fw}{{\mathfrak w}}

\newcommand{\sfc}{{\mathsf c}}

\newcommand{\sh}{{\mathsf h}}
\newcommand{\sll}{{\mathsf l}}

\newcommand{\so}{{\mathsf o}}
\newcommand{\sq}{{\mathsf q}}

\newcommand{\Loc}{{\SL\so\sfc}}

\newcommand{\homsl}{\CM^{\BC}_{\rm flat}(C)}
\newcommand{\homslr}{\CM^{\BR}_{\rm flat}(C)}

\newcommand{\FV}{{\mathfrak V}}

\newcommand{\0}{{\mathfrak 0}}

\newcommand{\one}{{\mathfrak 1}}
\newcommand{\two}{{\mathfrak 2}}
\newcommand{\three}{{\mathfrak 3}}

\newcommand{\BF}{{\mathbb F}}

\newcommand{\BR}{{\mathbb R}}

\newcommand{\BC}{{\mathbb C}}

\newcommand{\BP}{{\mathbb P}}
\newcommand{\BS}{{\mathbb S}}

\newcommand{\BU}{{\mathbb U}}
\newcommand{\BZ}{{\mathbb Z}}

\newcommand{\rf}[1]{(\ref{#1})}

\newcommand{\Fus}[6]{F_{#5#6}^{}\big[\,{}^{#3}_{#4}\;{}^{#2}_{#1}\,\big]}

\newcommand{\nc}{\newcommand}

\nc{\rnc}{\renewcommand} \nc{\beq}{\begin{equation}}
\nc{\eeq}{\end{equation}} \nc{\beqa}{\begin{eqnarray}}
\nc{\eeqa}{\end{eqnarray}}

\begin{document}
\title{Supersymmetric gauge theories,
quantization of $\CM_{\rm flat}$, and conformal field theory}
\author{J. Teschner}
\address{
DESY Theory, Notkestr. 85, 22603 Hamburg, Germany}
\maketitle

\begin{quote}
\centerline{\bf Abstract}
{\small  We review the relations between 
$\CN=2$-supersymmetric gauge theories, Liouville theory
and the quantization of moduli spaces
of flat connections on Riemann surfaces.}\end{quote}

%\subsection{Definition}

%\tableofcontents

%\newpage

\section{Introduction}
\setcounter{equation}{0}

Alday, Gaiotto and Tachikawa \cite{AGT} discovered remarkable relations
between the instanton partition functions of certain four-dimensional 
$\CN=2$-supersymmetric gauge theories and the conformal field theory called Liouville
theory. These relations will be referred to as the  AGT-correspondence.
We will discuss an explanation for the AGT-correspondence
based on the observation that both instanton partition functions
and Liouville conformal blocks are naturally related to 
certain wave-functions in the quantum theory obtained by quantising
the moduli spaces of flat ${\rm PSL}(2,\BR)$-connections on 
certain Riemann surfaces $C$. We will be considering a  
class of gauge theories referred to as class $\CS$, see 
\cite{G,GMN2} or the contribution \cite{G13} in this volume. The gauge theories
$\CG_{C,\mathfrak g}$ have elements labelled
by the choice of a Riemann surface $C$ and a Lie-algebra $\mathfrak g$ of type
$A$, $D$ or $E$. In the following we will restrict attention to the case where 
$\mathfrak g=A_1$, and denote the corresponding gauge theories as $\CG_C$.
However, the reader will notice that many of the arguments below generalise 
easily to more general theories of class $\CS$. 

The root for the relations between the gauge theories 
and moduli spaces of flat connections will be found in 
the identification of the algebra generated by
the supersymmetric  Wilson- and 't Hooft loop operators with the algebra of 
trace-functions which represent natural coordinates for the moduli spaces 
of flat connections. 
This algebra may become non-commutative
if the gauge theories are defined on curved spaces, 
or deformed by 
supersymmetry-preserving deformations like the Omega-deformation \cite{N}.
It turns out that the resulting non-commutativity
is the same as the one resulting from 
the quantisation of the relevant moduli 
spaces of flat connections.

Concerning the other side of the coin we are going to review the 
definition of the conformal blocks of Liouville theory. Formulated 
in the right way, part of the relation
to the quantization of moduli spaces of flat connections becomes 
obvious. There furthermore exists a natural
representation of the quantized algebra of trace functions
on the spaces of conformal blocks.

We are going to explain how the AGT-correspondence 
follows from the relation between
supersymmetric loop operators and trace functions, 
combined with certain consequences
of unbroken supersymmetry. Knowing precisely which 
algebra is generated by the  supersymmetric loop operators, 
% it turns out that 
one may reconstruct
expectation values of loop 
operators on backgrounds like the four-ellipsoid. From these data 
one may in particular
recover the low-energy effective actions of the considered
gauge theories. 
%It was therefore proposed in \cite{TV2} to call the algebra 
%of supersymmetric loop operators the "skeleton" of such gauge theories.
This approach relates
the AGT-correspondence to some of the work of Gaiotto, Moore and Neitzke
\cite{GMN2,GMN3}.  It is in some respects similar to the one
used by Nekrasov, Rosly and Shatashvili \cite{NRS} to study the
case with Omega-deformation preserving two-dimensional 
$\CN=2$ super-Poincar\'e invariance.

%There is an alternative approach towards proving the AGT-correspondence,
%which relates the series expansion 
%of $\CZ^{\rm inst}(a,m,\tau,\ep_\1,\ep_\2)$
%defined from the equivariant cohomology of
%instanton moduli spaces more directly to the 
%definition of the conformal blocks of Liouville theory obtained 
%from the representation theory of the Virasoro algebra \cite{AFLT}.

%The cases with genus $g>0$ 
%cause certain additional mathematical 
%headaches treated in \cite{TV2}.

\section{Theories of class $\CS$}

\setcounter{equation}{0}

\subsection{$A_1$ theories of class $\CS$.}

To a Riemann surface $C$ of genus $g$ and $n$ punctures 
one may associate \cite{G09,GMN2,G13} a four-dimensional gauge theory $\CG_C$ with 
$\CN=2$ supersymmetry, gauge group $({\rm SU}(2))^{h}$, $h:=3g-3+n$ 
and
flavor symmetry $({\rm SU}(2))^n$. The theories in this class are UV-finite, and therefore
characterised by a collection of gauge coupling {\it constants} $g_1,\dots,g_h$.
In the cases where $(g,n)=(0,4)$ and $(g,n)=(1,1)$ one would 
get the supersymmetric gauge theories commonly referred to
as $N_f=4$ and $\CN=2^*$-theory, respectively. The 
correspondence between data associated to the surface $C$ and 
the gauge theory $\CG_C$ is summarised in the table below.
\begin{center}
\begin{tabular}{l|l}
Riemann surface $C$ & Gauge theory $\CG_C$ \\ \hline\hline \\[-2ex]
Pants decomposition $\CC$ + trivalent 
 &  Lagrangian description with \\[1ex]
graph $\Gamma$ on $C,$ $\si=(\,\CC\,,\,\Ga\,)$ & action functional $S^\si_\tau$
\\[1ex] \hline \\[-2ex]
Gluing parameters $q_r=e^{2\pi i \tau_r}$, 
& UV-couplings $\tau=(\tau_1,\dots,\tau_h)$, \\
$r=1,\dots,3g-3+n$ & $\displaystyle{\tau_r=\frac{4\pi i}{g_r^2}+\frac{\theta_r}{2\pi}}$
\\[2ex] \hline \\[-2ex]
$r$-th tube & $r$-th vector multiplet $(A_{r,\mu},\phi_{{r}},\dots)$ \\[1ex]
$n$ boundaries  & $n$ hypermultiplets %, mass parameters $M_k$ 
\\[1ex] \hline \\[-2ex]
Change of pants decomposition & S-duality \\[2ex] \hline 
%\\[-2ex]
%Geodesic circumference of $r$-th tube & Coulomb branch modulus $a_r$
\end{tabular}
\end{center}

More details can be found in \cite{G13} and references therein.
To the $k$-th boundary 
there corresponds a flavor group $SU(2)_k$ with
mass parameter $M_k$. The hypermultiplet masses are linear 
combinations of the parameters $m_k$, $k=1,\dots,n$ as explained in more 
detail in \cite{G09,AGT}. 
The relevant definitions and results from Riemann surface theory 
are collected 
in Appendix \ref{RS}.
It is necessary to refine the pants 
decomposition by introducing 
the trivalent graph $\Gamma$ in order to have data that distinguish
action functionals with theta angles $\theta_r$ differing by multiples
of $2\pi$. This will be done such that 
\begin{equation}\label{S-Dehn}
S_{\tau+e_r}^{\si}\,=\,S^{\de_r.\si}_\tau\,,
\end{equation}
where $e_r$ is the unit vector with $r$-th component equal to one,
and $\de_r.\si$ denotes the action of the Dehn twist along the $r$-th
tube on $\si=(\CC,\Ga)$, which will map the graph $\Ga$ on $C$ to another
one.

\subsection{Realisation of S-duality}

Different Lagrangian descriptions of the theories $\CG_C$ 
are related by S-duality.
Two actions $S^{\si_1}_{\tau_1}$ and $S^{\si_2}_{\tau_2}$ 
describe different perturbative
expansions for one and the same theory. The respective perturbative
expansions will be valid  in the regimes where 
all coupling constants $g_{1,r}$ and $g_{2,r}$ are small.
To formulate the meaning of S-duality more precisely let us assume that there 
exists a non-perturbative definition of $\CG_C$ allowing us to 
define normalised 
expectation values of observables $\CO$ like $\langle\!\langle 
\,\CO\,\rangle\!\rangle_{\CG_{C_\tau}}^{}$ non-perturbatively as 
functions of $\tau$, a set of parameters for the complex structure on 
$C$.  S-duality holds if for each observable $\CO$  
there exist functionals 
$\CF^{\si_i}_\CO$ constructed using the fields in actions $S^{\si_i}_{\tau_i}$
together with choices of coupling constants $\tau_i=\tau_i(\tau)$, $i=1,2$,
such that
\begin{equation}
\big\langle\!\big\langle \,
\CO\,\big\rangle\!\big\rangle_{\CG_{C_\tau}}\,\asymp\,
\big\langle\!\big\langle \,
\CF_{\CO}^{\si_1}\,\big\rangle\!\big\rangle_{S^{\si_1}_{\tau_1}}
\quad{\rm and}\quad 
\big\langle\!\big\langle \,
\CO\,\big\rangle\!\big\rangle_{\CG_{C_\tau}}\,\asymp\,
\big\langle\!\big\langle \,
\CF_\CO^{\si_2}\,\big\rangle\!\big\rangle_{S^{\si_2}_{\tau_2}} \,,
\end{equation}
in the sense of equality of asymptotic expansions.

\begin{figure}[t]
\epsfxsize4.5cm
\centerline{\epsfbox{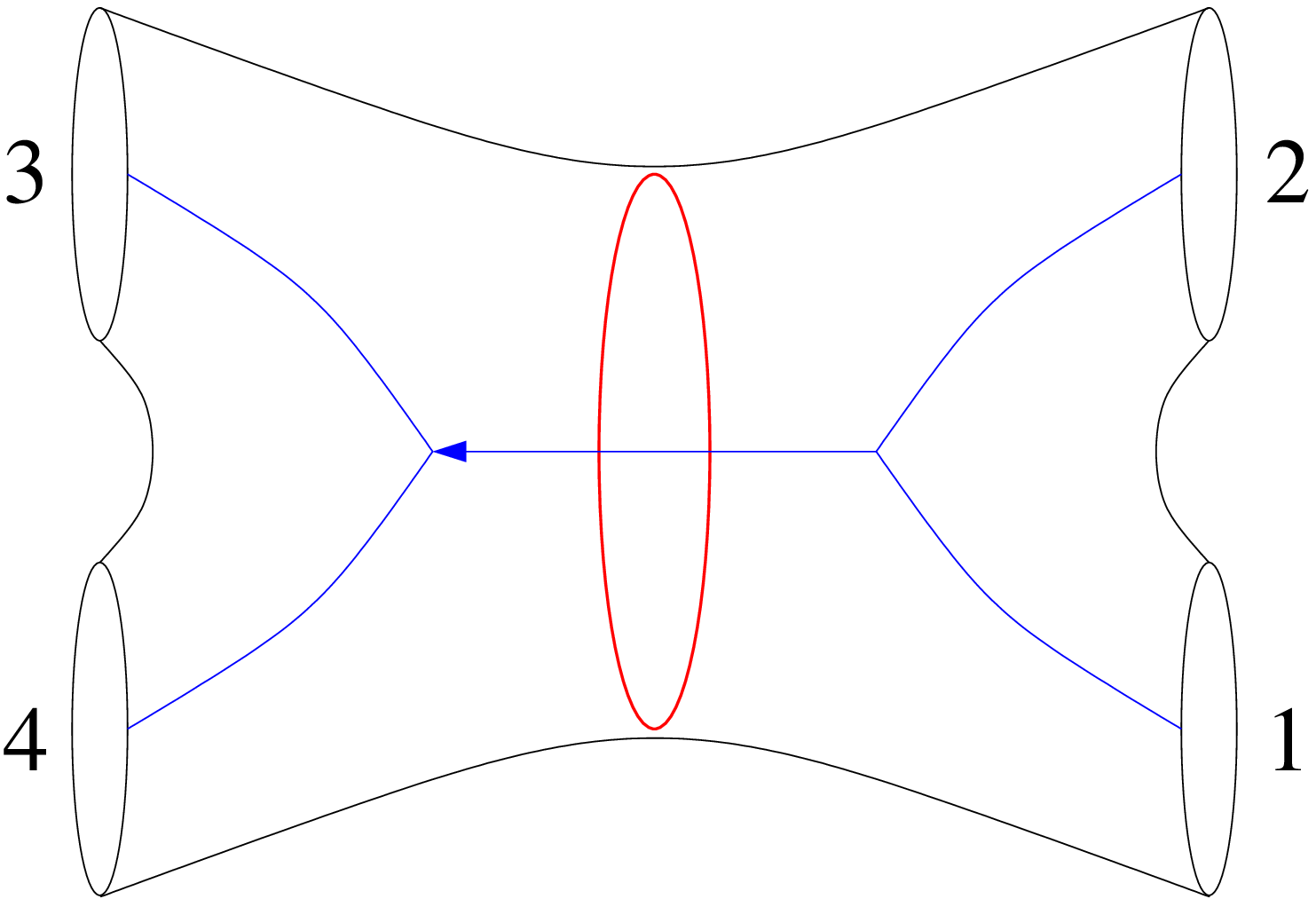}\hspace{.5cm}$\Longrightarrow$\hspace{.5cm}
\epsfxsize4.5cm\epsfbox{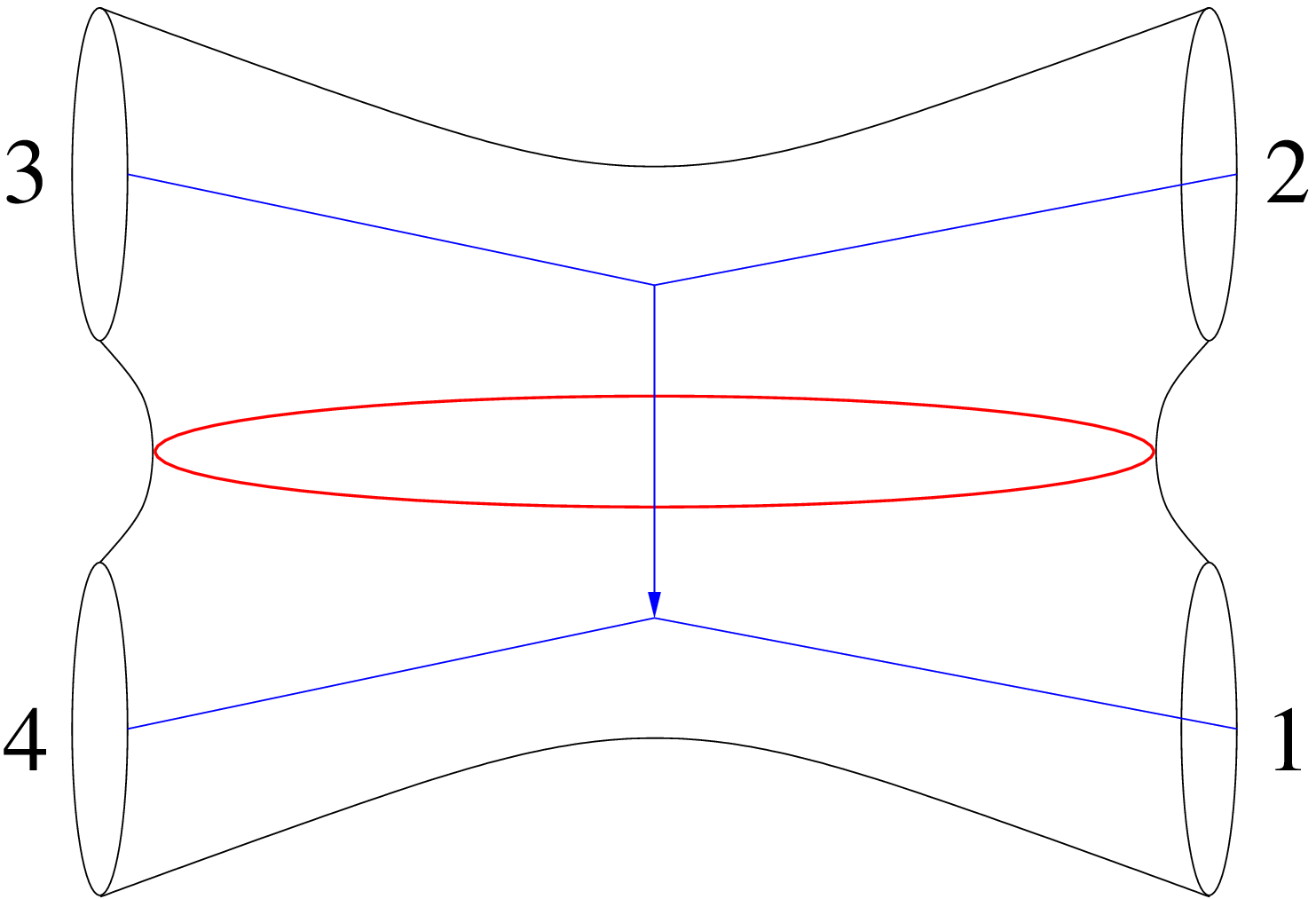}}
\caption{The F-move}\label{fmove}\vspace{.3cm}
\end{figure}
\begin{figure}[t]
\epsfxsize4.5cm
\centerline{\epsfbox{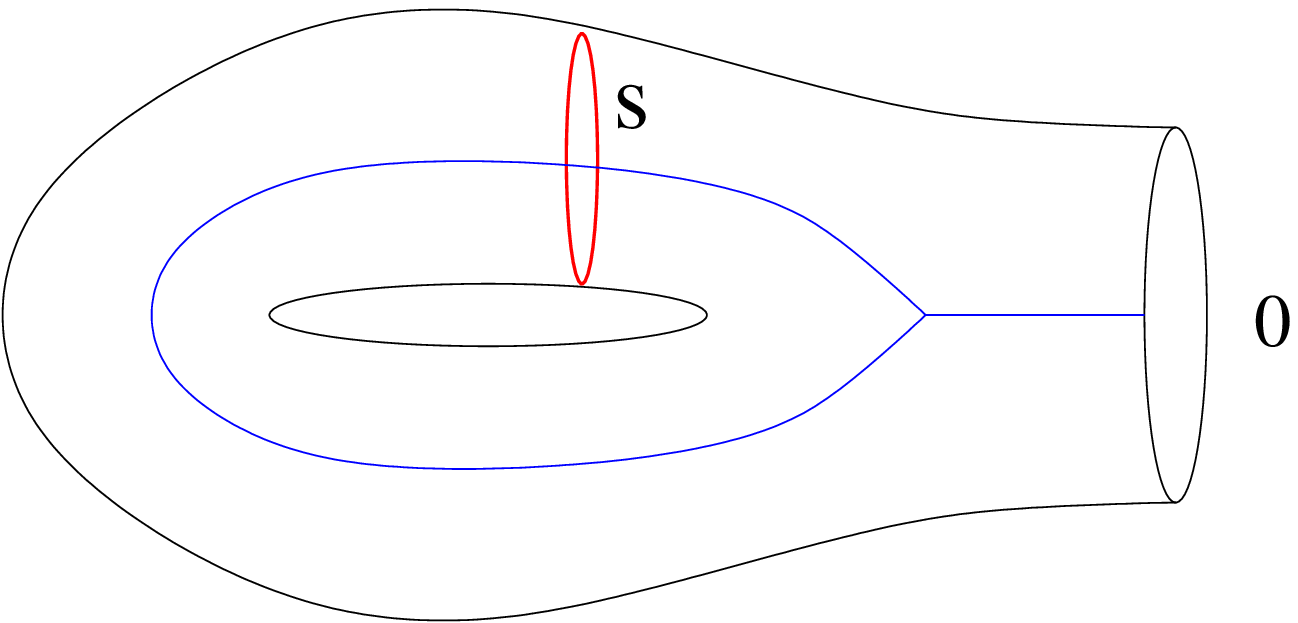}\hspace{.5cm}$\Longrightarrow$\hspace{.5cm}
\epsfxsize4.5cm\epsfbox{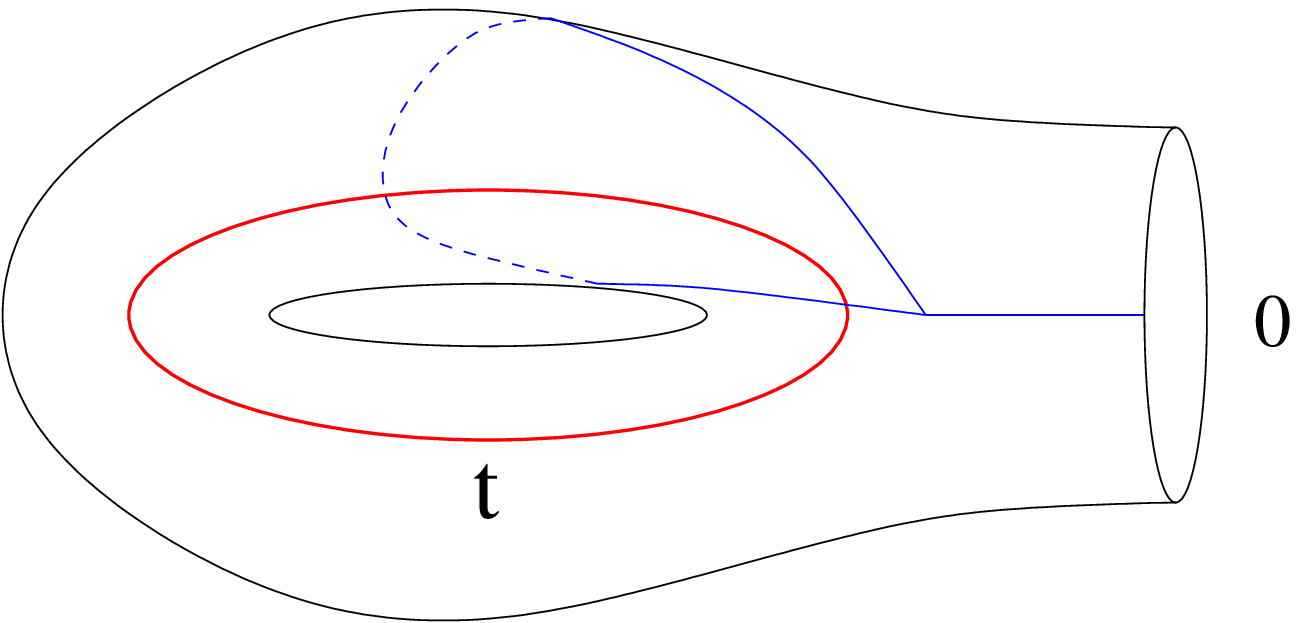}}
\caption{The S-move}\label{smove}\vspace{.3cm}
\end{figure}

\begin{figure}[htb]
\epsfxsize9cm
\centerline{\epsfbox{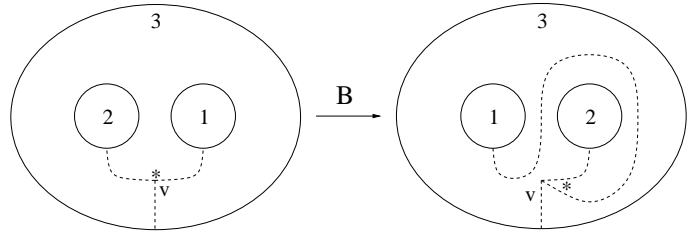}}
\caption{The B-move, represented by (indecently) looking into the pair of pants from above.}\label{bmove}\vspace{-.03cm}
\end{figure}

The passage from one Lagrangian description  $S^{\si}_{\tau}$ to another may be 
decomposed into the elementary S-duality transformations corresponding
to the cases where one
of the coupling constants $g_r$ gets large, while all others $g_s$, $s\neq r$
stay small.  The arguments given in \cite{G09} suggest that 
S-duality is realized in the following way:
In the regime where  $q_r=e^{2\pi i \tau_r}\ra 1$
one may use the Lagrangian description with action $S^{\si\!_{;r}}_{\tau'}$
associated to the data $\si\!{}_{;r}=(\,\CC_{;r}\,,\,\Ga\!_{;r}\,)$ obtained from 
$\si=(\CC,\Ga)$ by a local modification which is defined as 
follows: 
%To the curve $\ga_r\in\CC$
%let us associate the subsurface $C^r\hookrightarrow C$  which is the union of 
%three-holed spheres that have $\ga_r$
%in its boundary. $C^r$ may either be a four-holed sphere $C^r=C_{0,4}$ or a one-holed torus
%$C^r=C_{1,1}$. 
There is a unique subsurface 
$C_{r}\hookrightarrow C$
isomorphic to either $C_{0,4}$ or $C_{1,1}$ that contains $\ga_r$ in 
the interior of $C_r$.
$\si\!{}_{;r}=(\,\CC_{;r}\,,\,\Ga\!{}_{;r}\,)$ is defined by local substitutions within $C_r$
depicted in Figures \ref{fmove} and \ref{smove} for the two cases, respectively. If $C_r=C_{0,4}$ there is another strongly coupled regime which can be 
described in terms of a dual action. It corresponds to $q_r\ra\infty$, and the 
dual action $S^{\si\!{}_{:r}}_{\tau'}$ is associated to the data $\si\!{}_{:r}$ obtained
from $\si$ by the composition of the B-move depicted in Figure \ref{bmove}
with an F-move.

An important feature of the mapping between the respective sets of 
observables is that
the Wilson- and 't Hooft loops defined using $S^{\si}_{\tau}$ 
will correspond to the 
't Hooft and Wilson loops defined using $S^{\si\!_{;r}}_{\tau'}$, respectively.
This is the main feature we shall use in the following. 

Any transition between two pants decompositions $\si_1$ and 
$\si_2$ can be decomposed into the elementary F-, S-, and B-moves. 
It follows that
the groupoid of S-duality transformations coincides with the
Moore-Seiberg groupoid for the gauge theories of class $\CS$,
see Appendix \ref{MSdef}. % for a definition.

\subsection{Gauge theories $\CG_C$ on ellipsoids}

It may be extremely useful to 
study quantum field theories on compact Euclidean space-times
or on compact spaces rather than $\BR^4$. 
Physical quantities get finite size corrections which 
encode deep information on 
the quantum field theory we study. The zero modes of the
fields become dynamical, and have to be treated quantum-mechanically.

In the case of supersymmetric quantum field theories there are 
not many compact background space-times that allow us to preserve
part of the supersymmetry. A particularly interesting family 
of examples was studied in \cite{HH}, extending the seminal work
of Pestun \cite{Pe}. A review can be found in the Article
\cite{PV} in this volume. 

%\subsubsection{The four-dimensional ellipsoid}

Let us consider gauge theories $\CG_C$ on the four-dimensional ellipsoid
\begin{equation}
{\BBE}_{\ep_1,\ep_2}^4:=\,\{\,(x_0,\dots,x_4)\,|\,x_0^2+
\ep_1^2(x_1^2+x_2^2)+\ep_2^2(x_3^2+x_4^2)=1\,\}\,.
\end{equation}
It was shown in \cite{Pe,HH}, see also \cite{PV}, for some
examples of gauge theories $\CG_C$ that one of the supersymmetries  $Q$
is preserved on
 ${\BBE}_{\ep_1,\ep_2}^4$. It should be possible to 
generalize the proof of existence of an unbroken
supersymmetry $Q$ 
to all four-dimensional $\CN=2$ supersymmetric field theories with a Lagrangian description.

%\subsection{Partition functions on the ellipsoid}

Interesting physical quantities include the 
partition function $\CZ_{\CG_C}$, or more generally expectation values of
supersymmetric loop operators $\CL_\ga$
such as the Wilson- and 't Hooft loops. 
Such quantities are formally defined by the path integral over
all fields on ${\BBE}_{\ep_1,\ep_2}^4$. It was shown in a few
examples for gauge theories from class $\CS$ in \cite{Pe,HH}, 
reviewed in \cite{PV}, how to
evaluate this path integral by means of the localization 
technique. A variant of the localization argument was used to 
show that the integral over all fields actually reduces to 
an integral over the locus in field space where the 
scalars $\phi_r$ take constant {\it real} values $\phi_r
= {\rm diag}(a_r, -a_r)={\rm const}$, 
and all other fields vanish. 
This immediately implies that the path integral reduces 
to an ordinary integral over the variables $a_r$. It seems clear
that this argument can be 
generalized to all theories of class $\CS$ with a Lagrangian.

For some theories $\CG_C$ it was found in \cite{Pe} that the 
result of the localization calculation of the partition function
takes the form
\begin{equation}\label{ZE4}
Z_{{\BBE}_{\ep_1,\ep_2}^4}^{\CG_C}(m,\tau;\ep_1,\ep_2)
\,=\,\int d\mu(a)\;|\CZ^{\rm inst}(a,m,\tau;\ep_1,\ep_2)|^2\,.
\end{equation}
The main ingredients are the instanton partition function 
$\CZ^{\rm inst}(a,m,\tau;\ep_1,\ep_2)$
which depends on the zero modes $a=(a_1,\dots,a_h)$ of the scalar fields,
hypermultiplet mass parameters $m=(m_1,\dots,m_n)$, UV gauge coupling 
constants $\tau=(\tau_1,\dots,\tau_h)$, and two 
parameters $\ep_1,\ep_2$. 
The instanton partition functions can be defined 
as the partition function of the Omega-deformation of $\CG_C$ on $\BR^4$
\cite{N},
and may be calculated by means of the instanton calculus \cite{LNS,MNS1,MNS2,NS04}, 
as reviewed in \cite{Ta} in this volume.

It is expected that the form \rf{ZE4} will hold for arbitrary 
theories $\CG_C$, but the instanton partition function 
$\CZ^{\rm inst}(a,m,\tau;\ep_1,\ep_2)$ can only be calculated
for the cases where $C$ has genus $0$ or $1$, and the pants
decomposition is of linear or circular quiver type, respectively.

\subsection{Supersymmetric loop operators}

%\subsection{Definition of supersymmetric loop operators}

Supersymmetric Wilson loops can be defined as path-ordered exponentials of the
general form
\begin{subequations}
\label{Wilsondef}
\begin{align}
& W_{r,i}:={\rm Tr}\,\CP\exp\bigg[
\oint_{\CC}ds\;(i\dot{x}^\mu A_\mu^r+|\dot{x}|\phi^r)\bigg]\,.
%A_{e,\vf}-\frac{2}{\ep_1}\big(\,\phi_e\,\cos^2\rho/2+\bar\phi_e\,
%\sin^2\rho/2\,\big)\Big)\bigg)\,,\\
%& W_{e,\2}:={\rm Tr}\,\CP\exp\bigg(\int_{S^1_\2}d\chi\;\Big(
%A_{e,\chi}-\frac{2}{\ep_2}\big(\,\phi_e\,\cos^2\rho/2+\bar\phi_e\,
%\sin^2\rho/2\,\big)\Big)\bigg)\,.
\end{align}
\end{subequations}
The choice of contour $\CC$ is severely constrained by the requirement that the 
resulting observable is supersymmetric. Two possible choices for the four-manifold $M^4$ of interest 
are $M^4=\BR^3\times S^1$ and the four-ellipsoid. In the first case one may take a 
contour $\CC$ that wraps the $S^1$. For the case $M^4=E^4_{\ep_1\ep_2}$ it was 
shown in \cite{Pe,HH,GOP} that these observables 
are left invariant by the supersymmetry $Q$ preserved on 
${\BBE}_{\ep_1,\ep_2}^4$ if $\CC$ is one of the contours
$\CC_i$, $i=1,2$,
with $\CC_\1$ and $\CC_\2$  being the circles with constant
$(x_0,x_3,x_4)=0$ and $(x_0,x_1,x_2)=0$, respectively. Throughout this section we will assume
that $\CC$ is identified with one of the two $\CC_i$.

The 't Hooft loop observables $T_{r,i}$, $i=1,2$, 
can be defined semiclassically for vanishing theta-angles
$\theta=0$ 
by the boundary condition 
\begin{equation}
F_r\,\sim\,\frac{B_r}{4}\ep_{klm}\frac{x^k}{|\vec{x}|^3}dx^m\wedge dx^l\,,
\end{equation}
near the contour $\CC$. The coordinates $x^k$, $k=1,2,3$, are
local coordinates for the space transverse to $\CC_i$, 
and $B$ is an element of the Cartan subalgebra
of $SU(2)$. In order to get supersymmetric observables
one needs to have a corresponding singularity at $S^1_i$ 
for the scalar
fields $\phi_r$. For the details of the definition and the
generalization to $\theta\neq 0$ we
refer to \cite{GOP}.

Application of the localisation technique to the 
calculation of Wilson loop operators  \cite{Pe,HH}, see \cite{PV,O} for 
reviews, 
leads to results of the form
\begin{equation}\label{Pestun}
\big\langle \,
W_{r,i}\,\big\rangle_{\BBE_{\ep_1\ep_2}^4}
%\langle \,\CL_{e,i}\,\rangle_{\CG_C}^{{\BBE}_{\ep_1,\ep_2}}
=\int d\mu(a)\;|\CZ^{\rm inst}(a,m,\tau;\ep_1,\ep_2)|^2\,
2\cosh(2\pi a_r/\ep_i)\,,
\end{equation}
where $i=1,2$. 
A rather nontrivial extension of the method from
\cite{Pe} allows one to treat the case of 't Hooft loops \cite{GOP}
as well, see \cite{O} for a 
review.  The result is of the following form: 
\begin{equation}\label{GOP}
\big\langle \,
T_{r,i}\,\big\rangle_{{\BBE}_{\ep_1,\ep_2}^4}
%\langle \,\CL_{e,r}\,\rangle_{\CG_C}^{{\BBE}_{\ep_1,\ep_2}}
=\int d\mu(a)\;(\CZ_{\rm inst}(a,m,\tau;\ep_1,\ep_2))^*\,
\,\CD_{r,i}\CZ_{\rm inst}(a,m,\tau;\ep_1,\ep_2)\,,
\end{equation}
with $\CD_{r,i}$ being a difference operator acting only
on the variable $a_r$ of $\CZ_{\rm inst}(a,m,\tau;\ep_1,\ep_2)$, 
which has coefficients that 
depend on $a$, $m$ and $\ep_i$, in general.

\subsection{Relation to quantum Liouville theory}
%\label{Liousec}
\label{Cfbl-vert}

The authors of \cite{AGT} observed in some examples of theories 
from class $\CS$ that one has (up to inessential factors $\CZ^{\rm spur}(m,\tau;\ep_1,\ep_2)$)
an equality between the instanton partition functions and the 
conformal blocks $\CZ^{\rm Liou}(\be,\al,\tau;b)$ of
Liouville theory,
\begin{equation}\label{AGT}
\CZ^{\rm inst}(a,m,\tau;\ep_1,\ep_2)\,=\,\CZ^{\rm spur}(m,\tau;\ep_1,\ep_2)\,
\CZ^{\rm Liou}(\be,\al,q;b)\,,
\end{equation}
assuming a suitable dictionary between the variables involved. The "spurious" factor 
$\CZ^{\rm spur}(m,\tau;\ep_1,\ep_2)$ will turn out to be  inessential, dropping out of normalised expectation 
values
\begin{equation}
\big\langle\!\!\big\langle \,
\CL_{\ga}\,\big\rangle\!\!\big\rangle_{{\BBE}_{\ep_1,\ep_2}^4}:=
\big(\big\langle \,
1\,\big\rangle_{{\BBE}_{\ep_1,\ep_2}^4}\big)^{-1}\big\langle \,
\CL_{\ga}\,\big\rangle_{{\BBE}_{\ep_1,\ep_2}^4}\,,
\end{equation}
as follows easily from the general form of the results for
the expectation values 
quoted in \rf{genloop}, and is therefore called ``spurious''.

We'll now briefly review the
definition of the right hand side of  \rf{AGT} for the cases of Riemann surfaces $C$
of genus zero with $n$ punctures. The definition for Riemann surfaces 
$C$ of arbitrary genus is discussed in \cite{TV2}.

%\subsection{Virasoro conformal blocks}\label{Vircfbl}

\renewcommand{\fw}{w}
\renewcommand{\fv}{v}
\renewcommand{\fe}{e}
\newcommand{\vir}{{\rm Vir}_c}
\newcommand{\CFB}{{\SC\SB}}
\newcommand{\HFB}{{\SH\SC\SB}}
\newcommand{\SFB}{{\SS\SC\SB}}

%\subsection{Conformal blocks as expectation values of 
%chiral vertex operators}

The Virasoro algebra ${\rm Vir}_c$ has generators $L_n$,  $n\in\BZ$,
and relations
\begin{equation}\label{Vir}
[L_n,L_m] = (n-m)L_{n+m}+\frac{c}{12}n(n^2-1)\de_{n+m,0}.
\end{equation}

The relevant conformal blocks can be constructed using
chiral vertex operators. Let us use the notation 
$\De_\al:=\al(Q-\al)$, with $Q$ being a variable parameterising the value
$c$ of the central element in \rf{Vir} as $c=1+6Q^2$. We will  
denote the highest weight representation with weight $\De_\be$
by $\CV_\be$. A chiral
vertex operator is an operator 
$V_{\be_2\be_1}^\al(z):\CV_{\be_1}\ra\CV_{\be_2}$ that satisfies
the crucial intertwining property
\begin{equation}\label{Virinter}
[\,L_n\,,\,V_{\be_2\be_1}^\al(z)\,]\,=\,z^n(z\pa_z+(n+1)\De_{\al})
V_{\be_2\be_1}^\al(z)\,.
\end{equation}
The property \rf{Virinter} defines the operator $V_{\be_2\be_1}^\al(z)$
as a formal power series in $z^k$
uniquely up to multiplication with a complex number. 
The normalization freedom can be
parameterized by the number $N_{\be_2\be_1}^\al$ defined by
\begin{equation}
V_{\be_2\be_1}^\al(z)\,e_{\be_1}\,=\,
z^{\De_{\be_2}-\De_{\be_1}-\De_{\al}}\big[
N_{\be_2\be_1}^\al e_{\be_2}+\CO(z)\big]\,,
\end{equation}
where $e_\be$ is the highest weight vector of the representation $\CV_\be$.
A particularly useful choice for the normalization
factor  $N_{\be_2\be_1}^\al$ will be
\begin{equation}\label{Ndef}
N_{\be_2\be_1}^\al=\sqrt{C(\bar\al_3,\al_2,\al_1)}\,,
\end{equation}
where 
$\bar\al_3=Q-\al_3$, and $C(\al_3,\al_2,\al_1)$ is the 
%function defined as
%\begin{align}\label{ZZform}
%C(\alpha_1, & \alpha_2,\alpha_3)=\left[\pi\mu\gamma(b^2)b^{2-2b^2}
%\right]^{(Q-\sum_{i=1}^3\alpha_i)/b}\times\\
%& \times \frac{\Upsilon_0\Upsilon(2\alpha_1)\Upsilon(2\alpha_2)
%\Upsilon(2\alpha_3)}{
%\Upsilon(\alpha_1+\alpha_2+\alpha_3-Q)
%\Upsilon(\alpha_1+\alpha_2-\alpha_3)\Upsilon(\alpha_2+\alpha_3-\alpha_1)
%\Upsilon(\alpha_3+\alpha_1-\alpha_2)}.
%\notag\end{align}
%The
%expression on the right hand side of \rf{ZZform} is constructed
%out of the
%special function $\up(x)$ which is related to the Barnes
%double Gamma function $\Ga_b(x)$ 
%as $\up(x)=(\Ga_b(x)\Ga_b(Q-x))^{-1}$. The function
%$C(\alpha_1,\alpha_2,\alpha_3)$ is known to be the expression for the 
three-point function in Liouville theory.
An explicit formula for $C(\al_3,\al_2,\al_1)$ was 
conjectured in \cite{DO,ZZ}, and a derivation was subsequently 
presented in \cite{T01}. 

Using the invariant bilinear form $\langle\,.\,,\,.\,\rangle_\be^{}:
\CV_{\be}\ot\CV_{\be}\ra \BC$ one may then construct conformal blocks 
as matrix elements of products of chiral vertex operators such as
\begin{equation}\label{blockdef}
\CZ_{s}^{\rm\sst Liou}(\be,\al,q;b):=
\big\langle\, e_{\al_{n}}\,,\,
V_{\al_n\,,\,\be_{n-3}}^{\al_{n-1}}(z_{n-1})\,V_{\be_{n-3}\,,\,\be_{n-2}}^{\al_{n-2}}(z_{n-2})
\cdots V_{\be_1\al_1}^{\al_2}(z_2)\,e_{\al_{1}}
\,\big\rangle_{\al_n}^{}\,.\\
\end{equation}
The parameters $q=(q_1,\dots,q_{n-3})$ are given by the ratios
$q_r=z_{r+1}/z_{r+2}$, with $r=1,\dots,n-3$.
Equation \rf{blockdef} defines 
conformal blocks associated to particular pants decompositions of $C_{0,n}$.
In the case of $n=4$, for example, one gets the conformal blocks associated to the pants decomposition depicted 
on the left of Figure \ref{fmove}.

We may now state the dictionary between the variables appearing in the relation \rf{AGT}
between Liouville conformal blocks and the instanton partition functions of the corresponding gauge theories:
\begin{subequations}\label{paramid}
\begin{align}
& %b^2\,=\,\frac{\ep_1}{\ep_2}\,,\qquad
q_r=\frac{z_{r+1}}{z_{r+2}}=e^{2\pi i\tau_r}\,,\qquad
\be_r\,=\,\frac{Q}{2}+i\frac{a_r}{\hbar}\,,\qquad r=1,\dots,n-3\,,\\
&\al_k\,=\,\frac{Q}{2}+i\frac{M_k}{\hbar}\,,\quad k=1,\dots,n\,,\qquad
\hbar^2\,=\,\ep_1\ep_2\,.\qquad % \qquad Q:=b+b^{-1}\,,
\end{align}
\end{subequations}

In order to construct 
conformal blocks associated to general pants decompositions of surfaces $C_{0,n}$ of genus
zero let us introduce the 
descendants of a chiral vertex operator $V_{\be_2\be_1}^\al(z)$. The descendants may
be defined 
as the family of operators 
$V_{\be_2\be_1}^\al[v](z):\CV_{\be_1}\ra\CV_{\be_2}$
that satisfy 
\begin{equation}\begin{aligned}
&V_{\be_2\be_1}^\al[L_{-2}v](z)\,=\,:T(z)V_{\be_2\be_1}^\al[v](z):\,,\\
&V_{\be_2\be_1}^\al[L_{-1}v](z)\,=\,\pa_zV_{\be_2\be_1}^\al[v](z)\,,
\end{aligned}\qquad 
V_{\be_2\be_1}^\al[e_{\al}](z)\,=\,V_{\be_2\be_1}^\al(z)\,,
\end{equation}
where $:T(z)V_{\be_2\be_1}^\al[v]z:$ is defined
as 
\begin{equation}\label{normord}
:T(z)V_{\be_2\be_1}^\al[v](z):\,=\,
\sum_{n\leq -2}z^{-n-2}L_nV_{\be_2\be_1}^\al[v]z+
V_{\be_2\be_1}^\al[v](z)\sum_{n\geq -1}z^{-n-2}L_n\,.
\end{equation}
With the help of the descendants one has a new way to compose 
chiral vertex operators, allowing us, for example, to construct
conformal blocks on $C=\BP^1\setminus\{0,z_2,z_3,\infty\}$ as 
\begin{align}\label{altblockdef}
&\CZ_{t}^{\rm\sst Liou}(\be,\al,z;b):=
\big\langle\, e_{\al_{4}}\,,\,
V_{\al_4\al_1}^{\be}\big[\,V_{\be\al_2}^{\al_3}(z_3-z_2) e_{\al_2}\,\big](z_2)
\,e_{\al_{1}}\,\big\rangle_{\al_4}^{}\,.
\end{align}
This conformal block is associated to the pants decomposition on the right of Figure \ref{fmove}.
By considering arbitrary 
compositions of chiral vertex operators one may construct conformal blocks associated to 
arbitrary pants decompositions of a surface $C$ with genus zero and 
$n$ boundaries.

The relations \rf{AGT} have fully been proven \cite{AFLT} in the cases where the relevant
conformal blocks are of the from \rf{blockdef} corresponding to the 
so-called linear % and circular 
quiver gauge theories. It is 
not straightforward to generalise this 
proof to more general pants decompositions like those corresponding to 
conformal blocks of the form \rf{altblockdef}. The 
technical difficulties encountered for more general pants decompositions 
are considerable and not yet resolved in general, see \cite{HKS2} for partial 
results in this direction.

\section{Reduction to quantum mechanics}
\label{Sec:Hamilton}
\setcounter{equation}{0}

\subsection{Localization as reduction to zero mode quantum mechanics}

We may assign to the expectation values $\langle \CL\rangle$  of a loop observable $\CL$ 
an interpretation in terms of
expectation values of operators $\SL_{\sst\CL}$ which act
on the Hilbert space obtained
by canonical quantization of the gauge theory $\CG_C$ on
the space-time $\BR\times {\BBE}_{\ep_1,\ep_2}^3$, where
${\BBE}_{\ep_1,\ep_2}^3$ is the three-dimensional ellipsoid defined
as
\begin{equation}
{\BBE}_{\ep_1,\ep_2}^3:=\,\{\,(x_1,\dots,x_4)\,|\,
\ep_1^2(x_1^2+x_2^2)+\ep_2^2(x_3^2+x_4^2)=1\,\}\,.
\end{equation}
This is done by interpreting the coordinate $x_0$ for
${\BBE}_{\ep_1,\ep_2}^4$ as Euclidean time. Noting that 
${\BBE}_{\ep_1,\ep_2}^4$ looks near $x_0=0$ as 
$\BR\times {\BBE}_{\ep_1,\ep_2}^3$, we 
expect to be able to represent partition functions 
$\CZ_{\CG_C}({\BBE}_{\ep_1,\ep_2}^4)$ or expectation values
$\big\langle \,
\CL\,\big\rangle_{\CG_C^{}({\BBE}_{\ep_1,\ep_2}^4)}$
as matrix elements of states in the Hilbert space $\CH_{\CG_C}$ 
defined by
canonical quantization of $\CG_C$ on  
$\BR\times {\BBE}_{\ep_1,\ep_2}^3$.
More precisely
\begin{equation}
\CZ_{\CG_C}({\BBE}_{\ep_1,\ep_2}^4)
\,=\,\langle\,\tau\,|\,\tau\,\rangle\,,\qquad
\big\langle \,
\CL\,\big\rangle_{{\BBE}_{\ep_1,\ep_2}^4}
\,=\,\langle\,\tau\,|\,\SL_{\sst\CL}\,|\,\tau\,\rangle\,,
\end{equation}
where $\langle\,\tau\,|$ and $|\,\tau\,\rangle$ are the states 
created by performing the path integral over the 
upper/lower half-ellipsoid
\begin{equation}
{\BBE}_{\ep_1,\ep_2}^{4,\pm}:=\,\{\,(x_0,\dots,x_4)\,|\,x_0^2+
\ep_1^2(x_1^2+x_2^2)+\ep_2^2(x_3^2+x_4^2)=1\;,\;\pm x_0>0\,\}\,,
\end{equation}
respectively, and $\SL_{\sst\CL}$ is the operator that represents the 
observable $\CL$ within 
$\CH_{\CG_C}$.

%\subsubsection{Localization -- Interpretation 
%in the functional Schr\"odinger picture}

The form \rf{Pestun}, \rf{GOP} of the loop operator expectation values
is naturally interpreted in the Hamiltonian framework as follows.
In the functional Schroedinger picture one would represent the 
expectation values 
$\big\langle \,
\CL\,\big\rangle_{{\BBE}_{\ep_1,\ep_2}^4}$
schematically in the following form
\begin{equation}\label{funSchroe}
\big\langle \,
\CL\,\big\rangle_{{\BBE}_{\ep_1,\ep_2}^4}
\,=\,\int [\CD\Phi]\;(\Psi[\Phi])^*\,\SL_{\sst \CL}\Psi[\Phi]\,,
\end{equation}
the integral being extended over all field configuration on the 
three-ellipsoid ${\BBE}_{\ep_1,\ep_2}^3$ at $x_0=0$. The wave-functional
$\Psi[\Phi]$ is defined by means of the path integral 
over the lower half-ellipsoid
${\BBE}_{\ep_1,\ep_2}^{4,-}$ with Dirichlet-type boundary conditions
defined by the field configuration $\Phi$.  

The fact that the path integral localizes to the locus ${\SL\so\sfc}_C^{}$
defined by
constant values $\phi_r={\rm diag}(a_r.-a_r)={\rm const.}$ of the scalars
and zero values for all other fields \cite{Pe,HH,PV} implies that the path integral
in \rf{funSchroe} can be reduced to an ordinary integral of the 
form
\begin{equation}\label{funSchroe-loc}
\big\langle \,
\CL\,\big\rangle_{{\BBE}_{\ep_1,\ep_2}^4}
\,=\,\int da\;(\Psi_\tau(a))^*\,\pi_0(\SL_{\sst\CL})\Psi_\tau(a)\,,
\end{equation}
with  $\Psi_\tau(a)$ defined by means of the path integral 
over the lower half-ellipsoid
${\BBE}_{\ep_1,\ep_2}^{4,-}$ with Dirichlet boundary conditions
$\Phi\in\Loc_C$, $\phi_r={\rm diag}(a_r,-a_r)$, $r=1,\dots h$.
The Dirichlet boundary condition $\Phi\in\Loc_C$, $\phi_r=a_r$ is naturally
interpreted as defining a Hilbert subspace $\CH_0$ within $\CH_{\CG_C}$.
States in $\CH_0$ can, by definition, be represented by wave-functions
$\Psi(a)$, $a=(a_1,\dots,a_h)$. 
$\pi_0(\SL_{\sst\CL})$ is the projection of $\SL_{\sst\CL}$ to $\CH_0$.

Note that the boundary condition $\Phi\in\Loc_C$ preserves the supercharge $Q$ 
used in the localization calculations of \cite{Pe,HH,PV} --
that's just what defined the locus $\Loc_C$ in the first
place. We may therefore
use the arguments from \cite{Pe,HH} to identify 
the wave-functions $\Psi_\tau(a)$ in \rf{funSchroe-loc} with the 
instanton partition functions,
\begin{equation}\label{Psi-Z}
\Psi_{\tau}(a)\,=\,\CZ_{\rm inst}(a,m,\tau;\ep_1,\ep_2)\,.
\end{equation}
The form of the results for expectation values of loop observables
quoted in \rf{Pestun}, \rf{GOP} is thereby naturally explained.

\subsection{S-duality of expectation values}\label{loop-S-dual}

In each Lagrangian description 
with action $S^{\si}_\tau$ 
one will be able to express loop operator expectation values 
in the form
\begin{equation}\label{genloop}
\big\langle \,
\CL\,\big\rangle_{{\BBE}_{\ep_1,\ep_2}^4}^{S^\si_\tau}
\,=\,\int da\;(\Psi_\tau^\si(a))^*\,\CD\!{}_{\sst\CL}^{\;\si}\,\Psi_\tau^{\si}(a)\,,
\end{equation}
defining representations of the algebra $\CA_{\ep_\1\ep_\2}$ 
in terms of operators $\CD\!{}_{\sst\CL}^{\;\si}$.
The Wilson loops $W_{r,\1}$ 
and $W_{r,\2}$ act diagonally as
operators of multiplication by
$2\cosh(2\pi a_r/\ep_1)$ and $2\cosh(2\pi a_r/\ep_2)$, respectively.
The 't Hooft loops $T_{r,i}$ will be represented by 
difference operators denoted as $\CD_{r,i}^{\si}$.

In order for S-duality to hold, we need that the representations 
of the algebra of loop operators associated to any two pants decompositions $\si_1$ 
and $\si_2$ are unitarily equivalent.
This means in particular that the 
eigenfunctions $\Psi_\tau^{\si_1}(a)$ and $\Psi_\tau^{\si_2}(a)$
must be 
related by an integral transformations of the 
form\footnote{Considering theories $\CG_C$ 
associated to Riemann surfaces with genus $g>1$ one has to allow for an additional factor on the
right hand side of the
relation \rf{S-duality}. This is discussed in \cite{TV2}.}
\begin{equation}\label{S-duality}
\Psi_\tau^{\si_\2}(a_2)\,=\,
\int da_1\;K_{\si_\2\si_\1}(a_2,a_1)\,\Psi_\tau^{\si_1}(a_1)\,.
\end{equation} 
If $S^{\si}_{\tau}$ and $S^{\si'}_{\tau'}$ are two actions 
with $\tau$ and $\tau'$ differing only by shifts of 
the theta-angle $\theta_r\ra\theta_r+2\pi k_r$, it follows from \rf{S-Dehn} that we must have 
\begin{equation}
\Psi_{\tau'}^{\si}(a)\,=\,\Psi_{\tau}^{\si'}(a)\,,
\end{equation}
with $\tau'=\tau+k_re_r$. By using the transformations \rf{S-duality} one finds that
we must have 
\begin{equation}\label{Witten-effect}
\Psi_{\mu.\tau}^{\si}(a)\,=\,\Psi_{\tau}^{\mu.\si}(a)\,,
\end{equation}
for any Dehn twist $\mu\in{\rm MCG}(C)$. The notation 
$ \Psi_{\mu.\tau}^{\si}(a)$ on the left hand side denotes the analytic
continuation of $\Psi_{\tau}^{\si}(a)$ with respect to $\tau$ defined
by the element $\mu\in{\rm MCG}(C)$. 

By combining \rf{S-duality} and \rf{Witten-effect} we get
\begin{equation}\label{RH}
\Psi_{\mu.\tau}^{\si}(a_2)\,=\,
\int da_1\;K_{\mu.\si,\si}(a_2,a_1)\,\Psi_\tau^{\si}(a_1)\,.
\end{equation} 
Assuming that we know the kernels $K_{\mu.\si,\si}(a_2,a_1)$,
we would thereby get a Riemann-Hilbert type 
problem\footnote{The Riemann-Hilbert problem is often formulated as the problem to find vectors of multivalued analytic
functions on a punctured Riemann surface $C$ with given monodromy, a representation of $\pi_1(C)$ in $SL(N,\BC)$.
Our equation \rf{RH} generalises the Riemann-Hilbert problem in two ways: 
The Riemann surface $C$ is replaced by the moduli space $\CM(C)$ of complex structures on the surface $C$, and
the monodromy takes values in the group of unitary transformations of an infinite-dimensional 
Hilbert-space rather than $SL(N,\BC)$.}
for the wave-functions
$\Psi_\tau^{\si}(a_1)$. Equation \rf{RH} describes the effect of a 
monodromy in the gauge theory parameter space in terms of an
integral transformation with kernel $K_{\mu.\si,\si}(a_2,a_1)$.

Let's note, however, that the kernels $K_{\mu.\si,\si}(a_2,a_1)$ are by
no means arbitrary: They are strongly constrained by the fact that
\rf{S-duality} must intertwine the representations of the algebra
$\CA_{\ep_\1\ep_\2}$ defined by the actions $S^{\si_1}$ and $S^{\si_2}$,
respectively. Concretely, we must have, in particular, 
\begin{equation}\label{intertw}\begin{aligned}
&\overrightarrow{\CD_{r,i}^{\si_2}} \cdot K_{\si_\2\si_\1}(a_2,a_1)\,=\,
K_{\si_\2\si_\1}(a_2,a_1)\,2\cosh(2\pi a_{1,r}/\ep_i) \,,\\
&K_{\si_\2\si_\1}(a_2,a_1)\cdot\overleftarrow{\CD_{r,i}^{\si_1}}\,=\,
2\cosh(2\pi a_{2,r}/\ep_i)\,K_{\si_\2\si_\1}(a_2,a_1)\,,
\end{aligned}\end{equation}
expressing the fact that S-duality exchanges Wilson and 't Hooft loops.
The equations represent a system of difference equations that 
turns out to determine
$K_{\si_\2\si_\1}(a_2,a_1)$ uniquely up to normalization.
This means that the kernels are essentially
determined by the representation theory of the algebra
$\CA_{\ep_\1\ep_\2}$. 

We will in the following describe how 
to identify the algebra $\CA_{\ep_\1\ep_\2}$.
This information may then be used \cite{TV2} to 
determine the kernels $K_{\mu.\si,\si}(a_2,a_1)$ defining 
the Riemann-Hilbert problem \rf{RH}. Fixing the $\tau$-asymptotics 
by means of perturbative information one gets a Riemann-Hilbert problem 
which has an essentially unique solution, thereby 
characterizing the wave-functions 
$\Psi_{\tau}^{\si}(a)$ completely.

Keeping in mind \rf{Psi-Z} we conclude that the instanton partition functions 
$\CZ^{\rm inst}$
can be characterized % completely 
using the representation theory of $\CA_{\ep_1\ep_2}$.
Note that the prepotential $\CF$ giving the low-energy effective action of $\CG_C$
is recovered from $\CZ^{\rm inst}$ via 
$\CF=\lim_{\ep_1,\ep_2\ra 0}\ep_1\ep_2\log\CZ^{\rm inst}$.
This means 
that the low-energy effective action is encoded abstractly within the
algebra $\CA_{\ep_1\ep_2}$. These observations motivate why this algebra was called ``non-perturbative skeleton''
of $\CG_C$ in \cite{TV2}.

\section{The algebra of loop operators}
\setcounter{equation}{0}

In order to realise the program outlined at the 
end of the previous section 
it will be essential to know the algebra $\CA_{\ep_1\ep_2}$ precisely.
We are now going to explain how $\CA_{\ep_1\ep_2}$ is 
related to the non commutative algebra obtained by quantising the 
space of functions on the moduli space of flat ${\rm PSL}(2)$-connections.
This section is meant to give a guide to the literature on the known relations between 
the algebra generated by the supersymmetric Wilson- and 't Hooft loop operators
on the one hand, 
and the (quantised) algebra of functions on the moduli space of flat ${\rm PSL}(2)$-connections
on the other hand.

\subsection{The algebra of supersymmetric loop operators}

The algebra of gauge theory observables 
contains the supersymmetric Wilson- and 't Hooft loop observables.
The product of such loop operators will generate
further loop observables supported at the same loop $\CC$.
The generalizations
of Wilson- and 't Hooft loop operators $\CL_{\ga}$ that are generated in this
way describe the effect of inserting heavy ``dyonic'' probe particles, 
and can therefore be
labelled by pairs $\ga=(r,s)$ 
of electric and magnetic charge vectors, see \cite{DMO}, and
the article \cite{O} for a review. We will be interested in the 
algebra $\CA$
generated by polynomial functions of the loop operators.
%$\SW_{e,i}$ and $\ST_{e,i}$ by $\CA_{i}\equiv \CA_i(C)$,
%where $i=1,2$.

One should note that the labelling of loop operators
by charges is based on a given Lagrangian description of the theory. 
A particularly simple example for the dependence of the underlying 
Lagrangian is provided by the Witten-effect: Two actions $S_1$ and $S_2$ 
which differ only by a shift of the theta-angle $\theta_r$ by $2\pi$ will
define the same expectation values after proper identification of the loop
operators: A loop observable with charge $\ga_1$ 
defined by $S_1$ gets identified with the loop observable 
with charge $\ga_2$ defined by $S_2$ iff the magnetic charges coincide 
and the electric charges of 
$\ga_1$ and $\ga_2$ differ by certain multiples of the 
magnetic charges. A precise statement for the $A_1$ theories of class $\CS$
of interest here can be found in \cite{DMO}, see also \cite{O}.

It should also be remarked that the precise specification of 
a gauge theory of class $\CS$ depends on certain discrete topological data 
\cite{AST,Ta13} defining in particular the set of allowed 
charges for the line operators. This phenomenon is related to interesting subtleties showing up when 
the gauge theory $\CG_C$ is studied on more general four-manifolds, but 
it is not relevant for what is discussed in this article as we are
exclusively dealing with four manifolds having the topology of the four-sphere.

% We will  denote the
%algebra generated by all such supersymmetric loop operators by
%$\CA\equiv\CA_{\ep_\1\ep_\2}(C)\equiv \CA_{\1}\times\CA_{\2}$.
%This algebra turns out to play a crucial role for understanding 
%supersymmetric field theories of class $\CS$. 
%In the following we are going to
%review known relations between the algebras $\CA_{\ep_i}$ and the
%algebras of functions on certain moduli spaces of flat connections
%on the Riemann surface $C$. Having a more explicit understanding 
%of the algebras $\CA_{\ep_i}$ together with general expectations on the
%realization of the S-dualities in theories of class $\CS$ will allow us 
%even to reconstruct physical quantities like the expectation values of
%loop observables \cite{TV2}.

\subsection{UV versus IR loop operators}\label{UVvsIR}

It will be instructive to consider the four-ellipsoid $E^4_{\ep_1\ep_2}$ in the limit where $\ep_1=0$. 
In this case the four-ellipsoid ${\BBE}_{\ep_1,\ep_2}^4$ degenerates into
${\BBE}_{\ep_2}^2\times \BR^2$, where 
\begin{equation}
{\BBE}_{\ep_2}^2:=\,\{\,(x_0,\dots,x_2)\,|\,x_0^2+
\ep_2^2(x_3^2+x_4^2)=1\,\}\,.
\end{equation}
This implies that only
the Wilson- and 't Hooft loops wrapped on the remaining circle $\CC_2$ will 
remain.
%and it turns out that the algebra $\CA_{1}(C)$ generated by these
%observables will become a commutative algebra $\CA(C)$. 
We may still relate expectation values to matrix elements
by choosing $x_0$ as (Euclidean) time coordinate. Near the 
``equator'' $x_0=0$, the two-ellipsoid 
${\BBE}_{\ep_2}^2$ looks like
$\BR\times S^1$. One may expect that
studying the gauge theory $\CG_C$ on 
$\BR^2\times{\BBE}_{\ep_1}^2$ will allow us make contact with 
the work of Gaiotto, Moore and Neitzke \cite{GMN1}-\cite{GMN3},
who have studied the gauge theories
$\CG_C$ on the circle compactification $\BR^3\times S^1$. 
Aspects relevant for us are reviewed in \cite{Ne}.

Considering the theory $\CG_C$ on $\BR^3\times S^1$ 
at low energies, it was  argued in  \cite{GMN1} 
that $\CG_C$ becomes effectively
represented by a three-dimensional sigma model with hyperk\"ahler 
target space $\CM(C)$. This means in particular that the hyperk\"ahler
space $\CM(C)$ represents the moduli space of vacua of 
$\CG_C$ on $\BR^3\times S^1$.  

The supersymmetric Wilson- and 't Hooft loops supported on 
$S^1$ are called UV line operators in \cite{GMN3,Ne}.
Vacuum expectation values of these line operators\footnote{Comparing with
\cite{Ne} let us note that on $\BR^3\times S^1$ one may consider families 
of line operators preserving different supersymmetries, parameterised 
by a parameter $\zeta$ in \cite{Ne}. We here focus on the case $\zeta=1$
corresponding to the line operators studied on $E^4_{\ep_1\ep_2}$.
Let us furthermore note that
the label $\ga$ used 
for UV line operators here is used for IR line operators in \cite{Ne}.} 
\begin{equation}
L_\ga(m):=\langle\, \CL_{\ga}\,\rangle_m^{}\,,\qquad m\in\CM(C)\,,
\end{equation}
represent coordinate functions
on the moduli space of vacua of $\CG_C$ on $\BR^3\times S^1$. 
We see that the algebra $\CA$ of UV line operators must coincide with 
a (sub-)algebra of the algebra of functions on the moduli space of vacua
$\CM(C)$.

Other useful sets of coordinate functions for $\CM(C)$ have been defined in 
\cite{GMN1} using 
the effective low-energy description of $\CG_C$: 
They are denoted as $\CX_{\eta}(m)$, 
are labelled by the charge lattice $\Ga$, and represent Darboux coordinates
for the holomorphic symplectic structure $\Omega$ on $\CM(C)$.
The functions $\CX_{\eta}(m)$ have been interpreted in \cite{GMN3} 
as expectation values of IR line operators describing the effect 
of the insertion of a heavy dyonic source of charge $\eta$ 
into the low-energy effective field theory.

It has been argued in \cite{GMN3}, see also \cite{Ne}, 
that the expectation values $L_{\ga}(m)$
can be alternatively computed using the effective IR description of 
$\CG_C$ on $\BR^3\times S^1$, leading to a relation between
UV and IR line operators of the following
form:
\begin{equation}\label{UV-IR}
L_{\ga}(m)\,=\,\sum_{\eta\in\Ga}\overline{\underline{\Omega}}_{\ga,\eta}\,
\CX_{\eta}(m)\,.
\end{equation}
The positive-integer 
coefficients $\overline{\underline{\Omega}}_{\ga,\eta}$ have an 
interesting physical interpretation as an index counting certain BPS states
that exist in the presence of line defects \cite{GMN3}.

\subsection{Relation with moduli spaces of flat connections}

The following table summarizes known connections between the 
moduli spaces $\CM_{\rm flat}(C)$ 
of flat $SL(2)$-connections\footnote{This may be $SL(2,\BC)$- or
$SL(2,\BR)$-connections depending on the context, as will 
be discussed later.} on the surfaces $C$ and the moduli space of 
vacua $\CM(C)$ on $\BR^3\times S^1$:
\begin{center}
\begin{tabular}{l|l}
Riemann surface $C$ & Gauge theory $\CG_C$ \\ \hline\hline \\[-2ex]
Moduli space of flat connections $\CM_{\rm flat}(C)$ & 
Moduli space of vacua $\CM(C)$ on $\BR^3\times S^1$
\\[1ex] \hline \\[-2ex]
Trace functions $L_{\ga}$ on $\CM_{\rm flat}(C)$ & UV line operators $\CL_{\ga}$\\[1ex] 
\hline \\[-2ex]
Fock-Goncharov coordinates & IR line operators $\CX_{\eta}$ \\[1ex] \hline 
\end{tabular}
\end{center}
We have gathered the relevant definitions and results concerning 
$\CM_{\rm flat}(C)$ in Appendix \ref{Mflat}. The mapping 
between trace functions $L_\ga$ and UV line operators $\CL_\ga$ 
is defined by 
identifying  
the 
Dehn-Thurston parameters classifying closed loops on $C$
(see Subsection \ref{Dehn} for a short summary)
with the charge labels $\ga=(r,s)$ of the line operators \cite{DMO,O}.\footnote{The set of allowed 
charges  $\ga=(r,s)$ in a theory $\CG_C$ is generically smaller than the set of allowed Dehn-Thurston parameters \cite{AST,Ta13}.
This subtlety does not affect our discussions: For each allowed Dehn-Thurston parameter there {\it exists}
a choice of the extra discrete data specifying gauge theories $\CG_C$ such that the corresponding 
UV line operator $\CL_{\ga}$ can be defined within $\CG_C$.
Having determined the set of allowed charges in the duality frame
corresponding to a particular pants decomposition, one may figure out 
the allowed charges in any other duality frame by some simple rules.} 
The definition of the Fock-Goncharov coordinates for $\CM_{\rm flat}(C)$ is briefly 
reviewed in Appendix \ref{FGcoords}, and the relations to IR line operators 
are discussed in \cite{GMN2,GMN3}.

An argument in favor of the 
identification between $\CM_{\rm flat}(C)$ and $\CM(C)$
starts by 
considering the six-dimensional $(2,0)$ theory on 
$S^1\times \BR^3\times C$. Compactifying first on 
$C$ and then on $S^1$ gives the three-dimensional sigma model
with target space $\CM(C)$, as mentioned above. It may
alternatively be obtained from the six-dimensional theory by
first compactifying on $S^1$ followed by compactification on $C$. 
After compactifying on $S^1$ one would then find the maximally
supersymmetric five-dimensional super-Yang-Mills theory on 
$\BR\times\BR^2\times C$. Further compactification on $C$ yields
a nonlinear sigma-model with target being $\CM_{\rm Hit}(C)$, the
moduli space of solutions to Hitchin's self-duality equations 
using a variant of the argument presented in \cite{BJSV}.
More details and references can be found in \cite[Section 3.1]{GMN2}.
$\CM_{\rm Hit}(C)$ is a hyperk\"ahler space naturally related
to $\CM_{\rm flat}(C)$ in one of its hyperk\"ahler structures \cite{Hi,Ne}.

A way to find the identification between particular coordinate
functions on  $\CM_{\rm flat}(C)$ and the UV line operators
summarised in the table above was described in \cite[Section 7]{GMN3}. 

\subsection{Quantization}\label{quant-gauge}

An interesting generalization of the set-up considered in Subsection 
\ref{UVvsIR} (compactification on $S^1$) is obtained by imposing 
certain twisted boundary conditions with parameter $b$ 
along $S^1$ \cite{GMN3,IOT}.
The resulting deformation, denoted $\BR^3\times_b S^1$
of the background  $\BR^3\times S^1$
is related to the Omega-deformation,
and it can be used to model the residual effect of the curvature
in the vicinity of the circles $\CC_i$ on ${\BBE}_{\ep_1,\ep_2}^4$  
which represent the support of the loop operators \cite{O}. 

It has been argued in \cite{GMN3,IOT}, see also \cite{O}, that the
effect of the twisted boundary conditions is to deform the algebra
$\CA$ into a non-commutative algebra $\CA_{b}$. 
In the case of the $A_1$ theories
of class $\CS$ it was argued in \cite{GMN3} that the resulting 
algebra is nothing but the quantized algebra of functions on 
$\CM_{\rm flat}(C)$,
denoted ${\rm Fun}_\hbar(\CM_{\rm flat}(C))$, here with $\hbar=b^2$.
There should in particular exist a deformed version of the relation
\rf{UV-IR} between UV and IR line operators. The left hand side of this
relation, the deformed UV line operator, should be independent of the 
choice of coordinates that appear on the right hand side. As different 
sets of coordinates $\CX_\eta$ are related by (quantized-) cluster transformations,
it will suffice to figure out the quantum analog of \rf{UV-IR} for particular
triangulations. This is what was done in \cite[Section 11]{GMN3} for the $A_1$-case, leading to
the conclusion that the algebra generated by the deformed UV line operators 
is the quantisation of the the algebra of trace functions on $\CM_{\rm flat}(C)$ that
will be described in more detail in the following section.

Highly nontrivial support for this proposal has been given by
explicit calculations for some theories of class $\CS$ \cite{IOT,O}.
A rather different line of arguments leading to the same conclusion 
was proposed by Nekrasov and Witten in \cite{NW}.

\subsection{Back to the ellipsoid}

As mentioned above, one may expect that the twisted boundary conditions 
defining $\BR^3\times_b S^1$ would model the residual effect of the curvature
in the vicinity of the curves $\CC_i$ on  ${\BBE}_{\ep_1,\ep_2}^4$ \cite{IOT,O},
at least as far as the algebraic properties of loop operators 
are concerned. The comparison of the
results of localisation calculations on the two spaces (\cite{GOP} for $S^4$, 
and \cite{IOT} for $\BR^3\times_b S^1$) provides highly nontrivial 
quantitative evidence for this claim. In the case of  the four ellipsoid 
${\BBE}_{\ep_1,\ep_2}^4$ one thereby
expects to get a (twisted) product of two copies of 
${\rm Fun}_\hbar(\CM_{\rm flat}(C))$ associated to the two circles $\CC_i$
supporting supersymmetric loop observables.

However, 
there is a crucial difference between the cases of $\BR^3\times_b S^1$ 
and ${\BBE}_{\ep_1,\ep_2}^4$. In the case of $\BR^3\times_b S^1$
one will generically
get complex values for expectation values of loop observables which are
functions of the scalar expectation values at infinity, the holonomy of the 
gauge field around $S^1$ and
the complexified gauge coupling constants $\tau$. The precise relation
was given in \cite{IOT}.

In the case of 
${\BBE}_{\ep_1,\ep_2}^4$, on the contrary, one gets 
only real numbers larger than $2$ for the expectation values of Wilson loops
from the localisation calculations of \cite{Pe,HH}. By S-duality this
will imply that the 't Hooft loops will define positive self-adjoint 
operators on $\CH_0$ with the same spectrum. 
This means that the relevant moduli spaces to 
consider in this case will not be 
the moduli spaces $\CM_{\rm flat}^{\BC}(C)$ of flat 
${\rm PSL}(2,\BC)$-connections, but rather its real slice 
$\CM_{\rm flat}^{\BR}(C)$ defined by having real values bounded below by
$2$ for all 
trace coordinates. 

It is known that $\CM_{\rm flat}^{\BR}(C)$ breaks up
into finitely many disconnected components $\CM_{\rm flat}^{\BR,d}(C)$, $|d|=0,\dots,2g-2+n$,
and there exists a distinguished component $\CM_{\rm flat}^{\BR,0}(C)$
which has the necessary properties. This component is 
isomorphic to the Teichm\"uller spaces of Riemann surfaces 
\cite{Go,Hi} and therefore 
referred to as the Teichm\"uller component, see Appendix \ref{uniflat}.

The resulting situation is summarised in the table below.
\begin{center}
\begin{tabular}{l|l}
Riemann surface $C$ & Gauge theory $\CG_C$ \\ \hline\hline \\[-2ex]
Quantised algebras of functions 
&  Algebra $\CA_{b}$ generated by  \\[1ex]
${\rm Fun}_{b^2}^{}(\CM_{\rm flat}^{}(C))$ & 
Wilson- and 't Hooft loops on $\BR^{3}\times_b S^1$
\\[1ex] \hline\hline \\[-2ex]Quantized algebras of functions 
&  Algebra $\CA_{\ep_1\ep_2}$ generated by  \\[1ex]
${\rm Fun}_{b^2}(\CM_{\rm flat}^{\BR,0}(C))\tilde{\times} 
{\rm Fun}_{b^{-2}}(\CM_{\rm flat}^{\BR,0}(C))$ &  
Wilson- and 't Hooft loops on ${\BBE}_{\ep_1,\ep_2}^4$
\\[1ex] \hline 
\end{tabular}
\end{center}
The notation $\tilde{\times} $ indicates that the representatives of the factors commute 
only up to a sign, in general.

\section{Quantization
of moduli spaces of flat connections}\label{q-Mflat}
\renewcommand{\CL}{L}

\setcounter{equation}{0}

We now have the input we need to develop the program outlined in Subsection
\ref{loop-S-dual} - the reconstruction of instanton partition functions
from the algebra of loop operators.
In the rest of this section we shall briefly describe the quantization
of $\CM_{\rm flat}^{\BR,0}(C)$.

 \subsection{Quantization of the Fock-Goncharov coordinates}

The simplicity of the Poisson brackets of the Fock-Goncharov coordinates
makes part of the quantization quite simple. To each edge
$e$ of a triangulation $\frt$ of a Riemann surface $C_{g,n}$
associate a quantum operator $\SX^{\frt}_e$ corresponding to the
classical phase space function $\CX_e^\frt$. Canonical
quantization of the Poisson brackets (\ref{poisson}) yields
an algebra  $\CB_{\frt}$ with generators $\SX^{\frt}_e$ and
relations\vspace*{-3pt}
\begin{equation}\label{comm}
\SX^{\frt}_e,\SX^{\frt}_{e^\prime}=e^{2\pi i b^2\,n_{ee'}}\SX^{\frt}_{e'}\,\SX^{\frt}_e\,,
\end{equation}
where $n_{ee'}$ is the number of intersections of $e$ with $e'$, 
counted with a sign.

Note furthermore that the variables $\CX_e$ are positive
for the Teichm\"uller component. The scalar product of the
quantum theory should realize the phase space functions
$\CX_e$ as {\it positive} self-adjoint operators $\SX^{\frt}_e$. 
By choosing a polarization one may
define a Schr\"odinger type representations $\pi_\frt$
in terms of multiplication and
finite shift operators. It can be realized on suitable dense subspaces
of the Hilbert
space $\CH_{\frt}\simeq L^2(\BR^{3g-3+n})$.

There exists a family of automorphisms
which describe the relation between the quantized
variables associated to different triangulations \cite{F97,Ka1,CF}. If  
triangulation $\frt_e$ 
is obtained from $\frt$ by changing only the diagonal in the quadrangle
containing $e$, we have 
\begin{equation}\label{q-cluster}
\SX_{e'}^{\frt_e}\,=\,\left\{
\begin{aligned} &\SX_{e'}^{\frt}\,\prod_{a=1}^{|n_{e'e}|}
\big(1+e^{\pi i(2a-1)b^2}
(\SX_e^{\frt})^{-{\rm sgn}(n_{e'e})}\big)^{-{\rm sgn}(n_{e'e})} \;\;&{\rm if}\;\;e'\neq e\,,\\
&(\SX_e^{\frt})^{-1}\;\;&{\rm if}\;\;{e'= e}\,.
\end{aligned}\right.
\end{equation}
It follows that the quantum theory of $\CM_{\rm flat}^{\BR,0}(C)$ 
has the structure
of a quantum cluster algebra \cite{FG2}.

It is possible to construct  \cite{Ka1} unitary operators
$\ST_{\frt_1,\frt_2}$ that represent the quantum cluster
transformations \rf{q-cluster} in the sense that
\begin{equation}
 \SX_e^{\frt_2}\,=\,
 \ST_{\frt_1\frt_2}^{-1}\cdot \SX_e^{\frt_1}\cdot\ST_{\frt_1\frt_2}\,.
\end{equation}
The operators $\ST_{\frt_2,\frt_1}$ describe the change of
representation when passing from the quantum theory
associated to triangulation $\frt_1$ to the one associated
to $\frt_2$.
It follows that the 
resulting quantum theory does not depend on the
choice of a triangulation in an essential way. 

{\it As indicated in Section \ref{quant-gauge}, one may intepret the
coordinates $\CX_e^\tau$ as expectation values of IR line operators. 
The formula
\rf{comm} describes the quantum deformation
induced by the twisted boundary condition on $\BR^3\times_b S^1$,
and \rf{q-cluster} describes the behavior of the IR line operators
under (quantum-) wall-crossing \cite{GMN1,GMN3,Ne}.}

\subsection{Quantization of the trace functions}\label{algebra}

There is a simple algorithm (reviewed in Appendix \ref{trace-FG})
for calculating the trace functions in terms
of the variables $\CX_e^\frt$ leading to Laurent polynomials in the 
variables $\CX_e$ of the form
\begin{equation}\label{L-X}
\CL_\ga\,=\,\sum_{\nu\in\BF}C^{\frt}_\ga(\nu)\,
\prod_{e}\,(\CX_e^\frt)^{\frac{1}{2}\nu_e}\,,
\end{equation}
where the summation is taken over a finite set $\BF$ of
vectors $\nu\in\BZ^{3g-3+2n}$ with components $\nu_e$. 

{\it According to \cite{GMN3,Ne} one may interpret the 
trace functions as UV line operators. Formula \rf{L-X}
thereby becomes identified with \rf{UV-IR}.}

For curves $\ga$ having $C^{\frt}_\ga(\nu)\in\{0,1\}$ for 
all $\nu\in\BF$ it has turned out to be 
sufficient to replace $(\CX_e^\frt)^{\nu_e}$ in \rf{L-X} by 
$\exp(\sum_{e}\nu_e\log\SX_e^\frt)$ in order to define 
the quantum operator $\SL_\ga^\frt$
associated to a classical trace function $\CL_\ga$.
For other triangulations one may define $\SL_\ga^{\frt'}$
using
\begin{equation}
\label{SLtau-inter}
\SL_\ga^{\frt'}=
\ST_{\frt\frt'}^{-1}\cdot \SL_\ga^{\frt}\cdot\ST_{\frt\frt'}\,.
\end{equation}
It turns out that this is sufficient to define the operators 
$\SL_\ga^\frt$ in general \cite{T05}. It follows from \rf{SLtau-inter}
that we may regard the algebras of quantised trace functions 
generated by the operators $\SL_\ga^\frt$ as different
representations $\pi_\frt$ of an abstract algebra 
$\CA_b$ which does not depend on the
choice of a triangulation, 
$\SL^\frt_\ga\equiv\pi_{\frt}(L_{\ga})$ for $L_\ga\in\CA_b$.

The operators $\SL_\ga^\frt$ 
are positive self-adjoint with spectrum bounded from below by $2$, 
as follows from
the result of \cite{Ka4}. 
Two operators $\SL_{\ga_1}^\frt$ and $\SL_{\ga_2}^\frt$ 
commute if the intersection
of $\ga_1$ and $\ga_2$ is empty. 
It is therefore possible to diagonalise simultaneously the 
quantised trace functions associated to a maximal set of 
non-intersecting closed curves
defining a pants decomposition \cite{T05,TV2}.

\subsection{Representations associated to pants decompositions}

Mutual commutativity of the quantized trace-functions $\SL_{\ga_r}^\frt$ 
ensures existence of operators $\SR_{\si|\frt}$ which map
the operators $\SL_{\ga_r}^\frt$, $r=1,\dots,h$ 
associated to the curves $\CC=(\ga_1,\dots,\ga_h)$ 
defining a pants decomposition to the 
operators of multiplication by  $2\cosh(l_r/2)$. 
The states in the image $\CH_\si$ of $\SR_{\si|\frt}$
can be represented by functions $\psi(l)$, 
$l=(l_1,\dots,l_h)$
depending on variables $l_r\in\BR^+$ which parameterise the
eigenvalues of $\SL_{\ga_r}^\frt$.
The operators $\SR_{\si|\frt}$ define a new family of 
representations $\pi_\si$ of $\CA_b$ via
\begin{equation}
\label{SLtau-sigma}
\pi_\si(L_\ga):=\SR_{\si|\frt}
\cdot \pi_\frt(L_\ga)\cdot(\SR_{\si|\frt})^{-1}\,.
\end{equation}
The representations are naturally labelled by the 
data $\si=(\,\CC\,,\,\Ga\,)$ we had encountered before.
The unitary operators $\SR_{\si|\frt}:\CH_\frt\ra\CH_\si$ 
were constructed explicitly
in \cite{T05,TV2}.

\subsubsection{Transitions between representation}

 The passage between the representations $\pi_{\si_1}$ 
and $\pi_{\si_2}$ associated to 
two different pants decompositions is then described by
\[
\SU_{\si_2\si_1}:=\SR_{\si_2|\frt}\cdot(\SR_{\si_1|\frt})^{-1}\,.
\]
The unitary operators $\SU_{\si_2\si_1}$ intertwine the
representations $\pi_{\si_1}$ and $\pi_{\si_2}$,
\begin{equation}\label{inter}
\pi_{\si_2}(\CL_\ga)\cdot \SU_{\si_2\si_1}\,=\,
\SU_{\si_2\si_1}\cdot \pi_{\si_1}(\CL_\ga)\,.
\end{equation}
Explicit representations for the operators 
$\SU_{\si_2\si_1}$ have been calculated in \cite{NT,TV2} 
for pairs $[\si_2,\si_1]$ related by the generators 
of the 
Moore-Seiberg groupoid. The B-move is represented as 
\begin{equation}\label{Bcoeff}
(\SB\psi_s)(\be)\,=\,B_{l_\2l_\1}^{l_3}\psi_s(\be)\,,
\qquad B_{l_\2l_\1}^{l_\3}\,=\,
e^{\pi i(\De_{l_\3}-\De_{l_\2}-\De_{l_\1})}\,,
\end{equation}
where $\De_l=(1+b^2)/4b+(l/4\pi b)^2$.
The F-move is represented in terms of an 
integral transformation of the form
\begin{equation}
\psi_s(l_s)\,\equiv\,(\SF\psi_t)(l_s)\,=\,
\int_{\BR^+} dl_t \;\Fus{l_1}{l_2}{l_3}{l_4}{l_s}{l_t}\;
\psi_t(l_t)\,.
\end{equation}
A similar formula exists for the S-move.
The explicit expressions can be found in \cite{TV2}.

The 
operators $\SU_{\si_\2\si_1}$ define 
a unitary projective 
representation of the Moore-Seiberg
groupoid, 
\begin{equation}\label{MS-abstr}
\SU_{\si_\3\si_2}\cdot\SU_{\si_\2\si_1}\,=\,\zeta_{\si_3\si_2\si_1}
\SU_{\si_\3\si_1}\,,
\end{equation}
where $\zeta_{\si_3\si_2\si_1}\in\BC$, $|\zeta_{\si_3\si_2\si_1}|=1$.
The explicit formulae for the relations of the 
Moore-Seiberg groupoid in the quantisation of $\CM^0_{\rm flat}(C)$ 
are listed in \cite{TV2}.

Having a representation of the Moore-Seiberg groupoid automatically
produces a representation of the mapping class group. 
An element of the mapping class group $\mu$ represents a diffeomorphism
of the surface $C$, and therefore maps any MS graph $\si$ 
to another one denoted $\mu.\si$. Note that the 
Hilbert spaces $\CH_\si$ and $\CH_{\mu.\si}$ are canonically 
isomorphic. Indeed, the Hilbert spaces $\SH_\si$ depend only on the combinatorics
of the graphs $\si$, but not on their embedding into $C$. We may therefore
define an operator $\SM_\si(\mu):\CH_\si\ra\CH_\si$ as
\begin{equation}\label{MCGdef}
\SM_\si(\mu):=\,\SU_{\mu.\si,\si}\,.
\end{equation}
It is automatic that the operators $\SM(\mu)$ define a projective 
unitary representation of the mapping class group ${\rm MCG}(C)$ 
on $\CH_\si$. 

{\it The kernels of the operators $\SU_{\si_2\si_1}$, $\ST_{\frt_2\frt_1}$
and $\SR_{\si |\frt}$ are
related to the partition functions of $d=3$ gauge theories on duality walls,
see \cite{DGV},\cite{D} and references therein. 
The relations are summarised in the
following table:} 
\begin{center}
\begin{tabular}{l|l}
Riemann surface $C$ & Gauge theory $\CG_C$ \\ \hline\hline \\[-2ex]
Kernels representing operators $\SU_{\si_2\si_1}$ 
& UV duality walls $T_2[M,{\mathbf p},{\mathbf p'}]$
\\[1ex] \hline \\[-2ex]
Kernels representing operators $\SR_{\si |\frt}$ & RG domain walls $T_2[M,{\mathbf p},\Pi]$
\\[1ex] \hline \\[-2ex]
Kernels representing operators $\ST_{\frt_2\frt_1}$ & IR duality walls 
$T_2[M,\Pi,\Pi']$
\\[1ex] \hline 
\end{tabular}
\end{center}

\subsubsection{Representations}\label{reprs}

The representations $\pi_\si(\SL_{\ga})$ were calculated 
explicitly for the
generators of $\CA_b$ in \cite{TV2}.  

As a prototypical example let us consider the case where
$\si$ corresponds to the pants 
decomposition of $C_{0,4}$ depicted on the left of Figure \ref{fmove}.
We may associate generators 
$L_s$, $L_t$ and $L_u$ of $\CA_b$
to
the simple closed curves $\ga_s$, $\ga_t$, and $\ga_u$ 
introduced in Subsection \ref{sec:Genrel}, respectively. 
The generators $\CL_r$ $r=1,\dots,4$ are associated to the 
boundary components of $C\simeq C_{0,4}$. The representation 
of $\CA_b$  will be generated
from the operators $\SL_s$, $\SL_t$ and $\SL_u$ defined as
follows:
\begin{subequations}\label{rep0,4}
\begin{align}
\SL_s:=\,& 2\cosh(\sll/2)\,.\\
\SL_t:=\,&\frac{1}{ 2(\cosh \sll_s - \cos 2\pi b^2)}
\Big(2\cos\pi b^2(L_2L_3+L_1L_4)+
\SL_s (L_1L_3+L_2L_4)\Big) \\ \label{quantum't Hooft}
& \quad +  \sum_{\ep=\pm 1}
\frac{1}{\sqrt{2\sinh(\sll_{s}/2)}}
e^{\ep\sk/2}
\frac{\sqrt{c_{12}(\SL_s)c_{34}(\SL_s)}}{2\sinh(\sll_s/2)}
e^{\ep\sk/2}
\frac{1}{\sqrt{2\sinh(\sll_s/2)}}  
 %[-.5ex]
%& \quad +  \frac{1}{\sqrt{2\sinh(\sll_s/2)}}
%{e^{-\sk_s/2}}
%\frac{\sqrt{c_{12}(\SL_s)c_{34}(\SL_s)}}{2\sinh(\sll_s/2)}
%{e^{-\sk_s/2}}
%\frac{1}{\sqrt{2\sinh(\sll_s/2)}}\,, 
\nonumber 
\end{align}
where 
\[
\sll\,\psi_\si(l)\,=\,
l_s \psi_\si(l)\,,\qquad
\sk\,\psi_\si(l)\,=\,-4\pi{\mathrm i} b^2\,\pa_{l}\psi_\si(l)\,,
\]
and  $c_{ij}(L_s)$ is defined as
\begin{align}\label{cijdef}
c_{ij}(L_s) & \,=\,L_s^2+L_i^2+L_j^2+L_sL_iL_j-4 \,.
%\\ \nonumber & =
%2 \cosh \fr{l_s+l_i+l_j}{4} 2 \cosh \fr{l_s+l_i-l_j}{4}
%2 \cosh \fr{l_s-l_i+l_j}{4} 2 \cosh \fr{l_s-l_i-l_j}{4}.
\end{align}
\end{subequations}
$\SL_u$ is given by a similar expression \cite{TV2}.
The operators $\sll_s$ and $\sk_s$ are quantum counterparts
of the Fenchel-Nielsen coordinates, see Appendix \ref{sec:FN} for a
definition.

{\it As indicated above, one may interpret the trace functions $L_s$, 
$L_t$, $L_u$ as UV line operators, here for the $N_f=4$ theory 
associated to $C_{0,4}$. $L_s$, $L_t$ and $L_u$ correspond to the 
Wilson loop, 't Hooft loop, and simplest dyonic loop, respectively. 
The formulae above are directly
related to the expectation values of these line operators on 
$\BR^3\times_b S^1$ calculated in \cite{IOT}.}

\subsubsection{The algebra of trace functions}

Using the explicit representations for the generators of $\CA_b$ 
obtained in \cite{TV2} it becomes straightforward to calculate the
relations that they satisfy. As a prototypical example, let us 
again consider the case $C=C_{0,4}$. There are two main relations:

Quadratic relation:
\begin{align} \label{CR}
\CQ(\CL_s, \CL_t, \CL_u):= & \,e^{\pi \textup{i} b^2} \CL_s \CL_t - 
e^{-\pi \textup{i} b^2} \CL_t \CL_s  \\ &
\,- (e^{2\pi \textup{i} b^2} - 
e^{-2\pi \textup{i} b^2}) \CL_u - 
(e^{\pi \textup{i} b^2} - e^{-\pi \textup{i} b^2}) (\CL_1\CL_3+\CL_2\CL_4)\,.
\notag
\end{align}
Cubic relation:
\begin{align}
\CP(\CL_s, & \CL_t, \CL_u) = \,-e^{\pi\textup{i} b^2} 
\CL_s \CL_t \CL_u \\ 
& + e^{2\pi \textup{i} b^2} \CL_s^2 + 
e^{-2 \pi \textup{i} b^2} \CL_t^2 + e^{2\pi \textup{i} b^2} \CL_u^2\nonumber \\ 
& + e^{\pi\textup{i} b^2} 
\CL_s (\CL_3\CL_4 + \CL_1\CL_2) + e^{- \pi \textup{i} b^2} 
\CL_t (\CL_2\CL_3 + \CL_1\CL_4) + e^{\pi\textup{i} b^2} 
\CL_u (\CL_1\CL_3 + \CL_2\CL_4)  \notag\\
& + 
\CL_1^2+\CL_2^2+\CL_3^2+\CL_4^2+\CL_1\CL_2\CL_3\CL_4 -\big(2\cos\pi b^2)^2 \,.
\nonumber
\end{align}
The generators $L_k$, $k=1,\dots,4$ are 
central elements in $\CA_b(C_{0,4})$, associated to the boundary components.
The quadratic relations represent the deformation of the 
Poisson bracket \rf{loopPB}, while the cubic relation
is a deformation of the relation \rf{algrel}.

\section{Relation to Liouville theory}\label{q-Mflat2}

\setcounter{equation}{0}

Having worked out the quantization of $\CM_{\rm flat}^{\BR,0}(C)$, we have 
determined the monodromy data we need to define 
the Riemann-Hilbert
type problem discussed in 
Section \ref{loop-S-dual}. 
In order to derive the AGT-correspondence along these lines
it remains to observe that the Liouville conformal
blocks provide solutions to this Riemann-Hilbert problem.

Our goal in this section is to explain 
why Liouville conformal blocks are the wave-functions
solving the Riemann-Hilbert problem \rf{RH}. 
To this aim we are going to explain that
 \[
 \boxed{\begin{aligned} &\text{Liouville theory is just 
 another way to represent the}\\
 & \text{quantum theory of $\CM_{\rm flat}^{\BR,0}(C)$
 defined in Section \ref{q-Mflat}.}
\end{aligned}
}
\]
The identification between conformal blocks and wave-functions
in the quantum theory of moduli spaces of flat connections
will follow naturally.

\subsection{Complex-analytic Darboux coordinates for 
$\CM_{\rm flat}^0(C)$}

Our explanations will be based on the fact that $\CM_{\rm flat}^{\BR,0}(C)$
is isomorphic to the Teichm\"uller space $\CT(C)$ (see Appendix 
\ref{uniflat}). This implies that there
exists an alternative
quantisation scheme using {\it holomorphic} coordinates for $\CT(C)$. 
We are going to explain that the quantum theory in the resulting 
quantisation scheme is naturally
related to conformal field theory.

For simplicity, we will here restrict attention to $C=C_{0,4}=\BP^1\setminus
\{z_1,z_2,z_3,z_4\}$.
We do not lose generality when we 
assume that $z_1=0$, $z_3=1$, $z_4=\infty$. The value of 
$q:=z_2$ defines a complex-analytic coordinate for the 
moduli space $\CM(C)$ of complex structures on $C$.
The Fuchsian group corresponding to the complex structure 
parameterized by a value of $q$ defines a flat 
${\rm PSL}(2,\BR)$-connection. 
We may therefore 
regard $q$ as a local coordinate for $\CM_{\rm flat}^0(C)$ which
is related to the Fenchel-Nielsen coordinates $(k,l)$ in a
very complicated way. The relation becomes reasonably simple only
in the limit $|q|\ra 0\Leftrightarrow l\ra 0$, where one has
\begin{equation}
\frac{l}{2\pi}\,\simeq\,\frac{\pi}{\log(1/|q|)}\,,\qquad 2\pi k\,\simeq\,\arg(q)\,,
\end{equation}
where the notation $\simeq$ indicates equality to leading order in this limit.

The complicated nature of the dependence of $q$ on the Darboux coordinates
$(k,l)$ is reflected in the fact that the Poisson structure on
$\CM_{\rm flat}^{\BR,0}(C)$ is represented in terms of $q$ in a much more
complicated way. A useful way to describe the Poisson structure
using the coordinate $q$ is to find
a function $h=h(q,\bar q)$ that is canonically 
conjugate to $q$ in the sense
that \begin{equation}
\{\,q\,,\,h(q,\bar q)\,\}\,=\,-i\,.
\end{equation}
Such a function can be found from the metric $ds^2=e^{2\vf}dyd\bar y$
of constant negative curvature
associated to $q$ by writing the function $t(y)=-(\pa_y\vf)^2+\pa_y^2\vf$
in the form 
\begin{equation}\label{fouroper}
t(y)=\frac{\de_3}{(y-1)^2}+\frac{\de_1}{y^2}+\frac{\de_2}{(y-q)^2}+\frac{\upsilon}{y(y-1)}+
\frac{q(q-1)}{y(y-1)}\frac{h}{y-q}.
\end{equation}
The residue $h=h(q,\bar{q})$ in \rf{fouroper} is indeed the 
sought-for conjugate variable to $q$, as follows from the beautiful
results \cite{TZ87a,CMS,TZ03} that 
the classical 
Liouville action $S_{\rm cl}[\vf]$ is the K\"ahler potential for
the symplectic form on $\CT(C)\simeq \CM_{\rm flat}^0(C)$ corresponding
to the Poisson-structure we consider, and that
\begin{equation}
h(q,\bar{q})\,=\,-\frac{\pa}{\pa q}S_{\rm cl}[\vf]\,.
\end{equation}
The function $h(q,\bar q)$ is called the accessory parameter. 
Having real monodromy (subgroup of ${\rm PSL}(2,\BR)$) of the differential operator
$\pa_y^2+t$ clearly
requires fine-tuning of the residue $h$ in \rf{fouroper} in a way that depends on
the complex structure $q$.

\subsection{Quantization of complex-analytic Darboux coordinates for 
$\CM_{\rm flat}^0(C)$}

One may then consider an alternative representation for the 
quantum theory of $\CM_{\rm flat}^0(C)$ which is such that
the operator representing the complex-analytic coordinate $q$
is realized as a multiplication
operator $\sq$, 
\begin{equation}
\sq\,\psi(q)\,=\,q\psi(q)\,.
\end{equation}
The quantization of the observable $h$
should then give an operator $\sh$ that satisfies
\begin{equation}
[\,\sh\,,\,\sq\,]\,=\,b^2\,,
\end{equation}
and can therefore be represented as 
\begin{equation}
\sh\,\psi(q)\,=\,b^2\frac{\pa}{\pa q}\psi(q)\,.
\end{equation}
In order for such 
a representation to be equivalent to the representation we
had previously defined using the Darboux 
coordinates $(k,l)$ we should consider wave-functions $\phi(q)$ 
that are holomorphic in $q$.
Such a representation can be seen as an analog of the coherent
state representation of quantum mechanics.

It will be useful for us to think of 
the wave-functions $\psi(q)$ in such a representation as
overlaps $\langle \,q\,|\,\psi\,\rangle$ of the abstract state $|\,\psi\,\rangle$
with an eigenstate $\langle\,q\,|$ of the operator $\sq$.

\subsection{Geometric definition of the conformal blocks}

In order to see how the quantisation of $\CT(C)$ is related to conformal field theory,
let us present a more geometric approach to the definition of the conformal blocks going back to \cite{BPZ}.

Let $C$ be the Riemann surface $C=\BP^1\setminus\{z_1,\dots,z_n\}$ 
of genus 0 with $n$ marked points 
$z_1,\ldots,z_n$. At 
each of the marked points $z_r$, $r=1,\ldots,n$,  
let us choose the local coordinates 
$w_r=y-z_r$. 
We associate highest weight representations
$\CV_r$, of $\vir$ 
to $P_r$, $r=1,\ldots,n$. The representations
$\CV_r$ are generated from highest weight vectors $e_r$ with
weights $\De_r$. 

The conformal blocks
are then defined to be the linear functionals
$\CF:\CV_{[n]}\equiv\otimes_{r=1}^n\CV_r\ra\BC$ that
satisfy the invariance property
\begin{equation}\label{cfblvir}
\CF(T[\chi]\cdot v) = 0\quad \forall v\in\CV_{[n]},\quad \forall\chi\in
\FV_{{\rm out}},
\end{equation}
where $\FV_{{\rm out}}$ is the Lie algebra of meromorphic
differential operators on $C$ which may have poles only at
$z_1,\ldots,z_n$. The action of $T[\chi]$ on
$\otimes_{r=1}^n\CR_r\ra\BC$ is defined as
\begin{align}\label{Tdef}
T[\chi] = \sum_{r=1}^n {\rm id}\ot\dots\ot\underset{(\rm r-th)}{L[\chi^{(r)}]}\ot\dots\ot{\rm id},\quad
L[\chi^{(r)}] := \sum_{k\in\BZ} L_k^{} \chi_k^{(r)} \in \vir,
\end{align}
where $\chi_k^{(r)}$ are the coefficients of the 
Laurent expansions of $\chi$ at the points $P_1,\dots P_n$,
\begin{equation}\label{Laurent}
\chi(z_r) = \sum_{k\in\BZ} \chi_k^{(r)}\,w_r^{k+1} \,\pa_{w_r} \in
\BC(\!(w_r)\!)\pa_{w_r}\,, 
\end{equation}
with $\BC(\!(t)\!)$ being the space of Laurent series in the variable $t$.

The vector space
of conformal blocks associated to the Riemann surface $C$
with representations $\CV_r$ associated to the marked
points $P_r$, $r=1,\ldots,n$ will be denoted as
$\CFB(\CV_{[n]},C)$. It is the space of solutions to 
the defining invariance conditions \rf{cfblvir}.

The space $\CFB(\CV_{[n]},C)$ is infinite-dimensional in general. 
Considering 
the case $n=4$,  
for example, one may see this more explicitly
by noting that the defining
invariance property allows us to 
express the values $\CF(v_4\ot v_3\ot v_2\ot v_1)$ 
in terms of  the complex numbers
\begin{equation}\label{values}
\CZ^{(k)}(\CF,C):=\CF(e_4\ot e_3\ot L_{-1}^k e_2\ot e_1)\,,\qquad
k\in\BZ^{>0}\,,
\end{equation}
were $e_i$ are the highest weight vectors of $\CV_i$, $i=1,2,3,4$. 
We note
that $\CF$ is completely defined by the values $\CZ^{(k)}(\CF,C)$.
The space of conformal blocks
$\CFB(\CV_{[4]},C)$ is therefore isomorphic as a vector space to the 
space of {\it formal} power series in one variable.

This definition of conformal blocks is closely related, but not quite identical to the
one introduced previously in 
Section \ref{Cfbl-vert}. To indicate the relation let us note that matrix elements 
like  
\begin{equation} \label{cfbl-4}
\CF_\be(v_4\ot v_3\ot v_2\ot v_1):=
\big\langle\, e_{\al_{4}}\,,\,
V_{\al_4\,,\,\be}^{\al_{3}}(z_{3})\,V_{\be\,,\,\al_1}^{\al_{2}}(z_{2})
\,e_{\al_{1}}
\,\big\rangle_{\al_4}^{}\,,
%\big\langle\, v_{{4}}\,,\,
%V_{\al_4\al_1}^{\be}\big[\,V_{\be\al_2}^{\al_3}[v_3](z_3-z_2) v_{2}\,\big](z_2)
%\,v_{{1}}\,\big\rangle_{\al_4}^{}\,,
\end{equation}
will represent particular examples for conformal blocks as defined in this section.
Validity of the defining invariance property \rf{cfblvir} follows from  
the covariance properties of the chiral vertex operators. 

%It turns out, but this is highly
%nontrivial, that the conformal blocks $\CF_\be$, $\be\in Q/2+i\BR^+$
%defined in \rf{cfbl-4} generate a basis $\CFB(\CV_{[4]},C)$ 
%in a suitable sense.

\subsection{Deformations of the complex structure of 
$C$}

A key point that needs to be discussed about the spaces of
conformal blocks is the dependence on the complex structure
of $C$, here specified by the positions $z_1,\dots z_n$
of the marked points. 
There is a natural way to represent infinitesimal
variations of the complex structure of $C$ on the spaces of
conformal blocks. By combining the definition of conformal
blocks with the so-called ``Virasoro uniformization'' of
the moduli space ${\CM}_{0,n}$ of complex structures on
$C=C_{0,n}$ one may construct a natural representation of
infinitesimal motions on ${\CM}_{0,n}$ on the space of
conformal blocks.

The ``Virasoro uniformization'' of the moduli space
${\CM}_{0,n}$  may be formulated as the statement that the
tangent space $T{\CM}_{0,n}$ to ${\CM}_{0,n}$ at $C$ can be
identified with the double quotient
\begin{equation}\label{VirUni}
T{\CM}_{0,n} = \Ga(C\setminus\{z_1,\ldots,z_n\},\Theta_C)
\bigg\backslash \bigoplus_{k=1}^n \BC(\!(w_k)\!)\pa_k
\bigg/ \bigoplus_{k=1}^n w_k\BC[[w_k]]\pa_k,
\end{equation}
where  $\BC[[w_k]]$ are the spaces of Taylor series in the local coordinates $w_k$ for $k=1,\dots,n$,
respectively, and $\Ga(C\!\setminus\{z_1,\ldots,z_n\},\Theta_C)$ is the
space of vector fields that are holomorphic on
$C\setminus\{z_1,\ldots,z_n\}$, embedded into $\bigoplus_{k=1}^n \BC(\!(w_k)\!)\pa_k$ 
via \rf{Laurent}.

Given a tangent vector $\vartheta\in T{\CM}_{0,n}$, 
it follows from the Virasoro uniformization \rf{VirUni} 
that we may find elements
$\eta_{\vartheta}$ of $\bigoplus_{k=1}^n
\BC(\!(t_k)\!)\pa_k$, which represent $\vartheta$ via
\rf{VirUni}.
Let us then consider $\CF(T[\eta_\vartheta] v)$ with $T[\eta]$
being defined in \rf{Tdef} in the case that $v$  is the product of highest weight vectors, 
$v=e_n\ot\dots\ot e_1$. 
\rf{VirUni} allows us to define the
derivative $\de_{\vartheta} \CF(v)$ of $\CF(v)$ in the direction of 
 $\vartheta\in T{\CM}_{0,n}$ as
\begin{equation}\label{Viract}
\de_{\vartheta} \CF(v) := \CF(T[\eta_{\vartheta}^{}]v),
\end{equation}
 Dropping the condition that $\fv$ is a product of highest
weight vectors, one may use \rf{Viract} to define
$\de_{\vartheta} \CF$ in general. 
And indeed, it is well-known that \rf{Viract} 
leads to the definition of a canonical flat connection on 
the space $\CFB(\CV_{[n]},C)$ of conformal blocks \cite{BF}.

\subsection{Conformal blocks versus function on  $\CT_{0,n}$}

In the case $n=4$ it is easy to see that \rf{Viract} can be reduced
simply to 
\begin{equation}\label{Viract4}
\pa_z \CF(v_4\ot v_3\ot v_2\ot v_1)\,=\,
\CF(v_4\ot v_3\ot L_{-1}v_2\ot v_1)\,.
\end{equation}
Let us introduce the notation 
\begin{equation}\label{partdef}
\CZ^{\rm\sst Liou}(\CF,C)\,=\,\CF(e_1\ot\dots \ot e_n)\,,
\end{equation}
for the value of $\CF$ on the product of highest weight
vectors. 
Equation \rf{Viract4} allows us to  
identify the values $\CZ^{(k)}(\CF,C)$ 
defined in \rf{values} as 
the {\it k}-th derivatives of the
partition functions 
$\CZ^{\rm Liou}(\CF,C)$. We had seen above that the collection of the numbers 
$\CZ^{(k)}(\CF,C)$ characterizes the conformal blocks
$\CF\in \CFB(\CV_{[4]},C)$ completely.

One may define the parallel
transport of conformal blocks over $\CM_{0,4}$ via 
\begin{equation}\label{transport}
\CZ^{\rm Liou}(\CF,C_w)=\sum_{k=0}^{\infty}\frac{1}{n!}(w-z)^n
\CZ^{(k)}(\CF,C_z)\,.
\end{equation}
We see that there is a one-to-one correspondence between
functions $\CY(w)$ defined on some open, simply connected 
neighborhood $\CU_z$ of a point $z$ in $\CT_{0,4}$ and conformal 
blocks $\CF$ for which the series \rf{transport} converges for 
all $w\in\CU_z$: The Taylor expansion coefficients $\CY_k$
of $\CY(w)$ can be used to define a conformal block 
$\CF_\CY\in \CFB(\CV_{[4]},C)$ 
such that $\CZ^{(k)}(\CF_\CY,C)=\CY_k$.
Conversely, for ``well-behaved'' conformal blocks $\CF\in 
\CFB(\CV_{[4]},C_z)$
one may use \rf{transport} to define a family of
conformal blocks in a neighborhood $\CU_z(\CF)$. 

We are ultimately not interersted in the most crazy conformal blocks, 
but rather in those 
whose partition functions can be analytically continued over all of $\CT(C)$,
and which have reasonably mild singular behaviour at the boundaries of $\CT(C)$.
Such a subspace will be denoted $\CFB^{\rm reg}(\CV_{[4]},C)$.
It was proposed in \cite{TV2} that the conformal blocks defined previously with the help
of chiral vertex operators generate a basis for 
$\CFB^{\rm reg}(\CV_{[4]},C)$ in a suitable sense. 
This proposal is based on the highly nontrivial
results of \cite{T01,T03a} 
that the partition functions $\CZ^{\rm Liou}$ can be analytically 
continued over
all of $\CT(C)$, and that the bases associated to different pants 
decompositions are
linearly related.

\subsection{Verlinde loop operators}

The construction of conformal blocks using chiral vertex operators, or more generally
by gluing conformal blocks associated to three-punctured spheres gives another 
way to define a natural family of operators acting on spaces of conformal blocks. The resulting operators will be identified with quantized trace functions. 
We will describe the construction in the case of genus $0$ in terms of chiral 
vertex operators. 

Let us consider chiral vertex operators $V^{\al}_{\be_2\be_1}(z)$  
in the special case where $\al=-b/2$,
assuming that $Q$ is represented as $Q=b+b^{-1}$.
If furthermore $\be_2$ and $\be_1$ are related as
$\be_2=\be_1\mp b/2$, the vertex operators
$\psi_s(y)\equiv\psi_{\be_1,s}(y):=V^{-b/2}_{\be_1- sb/2,\be_1}(y)$, $s\in\{1,-1\}$,
are well-known to satisfy a differential equation
of the form
\begin{equation}\label{null}
\pa_y^2\psi_{\be_1,s}(y)+b^2:T(y)\psi_{\be_1,s}(y):=\,0\,,
\end{equation}
with normal ordering defined in \rf{normord}.
%\begin{equation}
%:T(y)\psi_{\be_1,s}(y):=\,\sum_{k<-1}y^{-k-2}L_k
%\psi_{\be_1,s}(y)+\psi_{\be_1,s}(y)\sum_{k\geq-1}y^{-k-2}L_k\,.
%\end{equation}
The chiral vertex operators
$\psi_{\be_1,s}(y)$ are called degenerate
fields. It follows from \rf{null} that matrix elements such as
\begin{align}\label{CFdef}
&\CF_{ss'}(\,\al;\be\,|\,z\,|\,y_0\,|\,y\,):=
\langle \,\al_4\,|\,\psi_{s}(y_0) \psi_{s'}(y) \,|\,\Theta_{s+s'}\,
\rangle\,,\\
&|\,\Theta_{s+s'}\,
\rangle:=
V_{\al_4+(s+s')\frac{b}{2}\,,\,\be}^{\al_{3}}(z_{3})
V_{\be\al_1}^{\al_2}(z_{2})V_{\al_1,0}^{\al_1}(z_1)
|\,0\,\rangle\,,
\notag\end{align}
will satisfy the partial differential equation $\CD_{\rm BPZ} \CF=0$,
with
\begin{equation}\label{BPZ}
\CD_{\rm BPZ}:=\, \frac{1}{b^2}\frac{\pa^2}{\pa y^2}+
\frac{\De_{-\frac{b}{2}}}{(y-y_0)^2}+\frac{1}{y-y_0}\frac{\pa}
{\pa y_0}+
\sum_{k=1}^{3}\left(\frac{\De_{\al_k}}{(y-z_k)^2}+\frac{1}{y-z_k}\frac{\pa}
{\pa z_k}\right)\,.
\end{equation}
As explained previously, we may regard
the matrix elements \rf{CFdef} as the partition functions of 
conformal blocks in 
$\CFB(\CV_{[6]}',C_{0,6})$, where now
$\CV_{[6]}'=\CV_{[4]}\ot\CV_{-b/2}^{\ot 2}.$

Using these ingredients it is straightforward to show that the
analytic continuation of the matrix elements
$\CF_{ss'}(\,\al;\be\,|\,z\,|\,y_0\,|\,y\,)$, $s,s'\in\{1,-1\}$ 
with respect to $y$ along closed
paths $\ga$ on $C_{0,{5}}$ can be expressed as a linear
combination of the matrix elements
$\CF_{ss''}(\,\al;\be'\,|\,z\,|\,y_0\,|\,y\,)$
having parameters $\be'$ that differ from $\be$ by integer
multiples of the parameter $b$,
\begin{equation}
\CF_{ss'}(\,\al;\be\,|\,z\,|\,y_0\,|\,y\,)\,=\,
\sum_{s''=\pm}\SM_\ga(\be,\ST_\be)_{ss'}^{s''}\cdot\CF_{ss''}(\,\al;\be\,|\,z\,|\,y_0\,|\,y\,)
\end{equation}
where ${\ST}$ is the
operator which 
shifts the argument $\be$ of $\CF_{ss'}$ by the amount $b$. The matrices 
$\SM_\ga$ define representations of the fundamental group $\pi_{1}(C_{0,5})$ 
on the space $\CFB(\CV_{[6]}',C_{0,6})$.

The definition of the Verlinde loop operators
is based on the simple fact that $\CFB(\CV_{[4]}\ot\CV_0,C_{0,5})$ 
is canonically isomorphic to $\CFB(\CV_{[4]},C_{0,4})$ 
if $\CV_0$ is the vacuum representation. 
One may furthermore note that there exists a 
linear combination $\sum_sK_s\psi_{s}(y_0) \psi_{-s}(y)$ 
which is a descendant of
the chiral vertex operator  
$V^{0}_{\be,\be}\big[\psi_+(y_0-y)e_{-\frac{b}{2}}\big](y)$ 
associated to the 
vacuum representation.
These observations allow us to define both an embedding $\imath$
and a projection $\wp$,
\begin{equation}
\begin{aligned}
&\imath:\CFB^{\rm reg}(\CV_{[4]},C_{0,{4}})\hookrightarrow \CFB^{\rm reg}(\CV_{[6]}',C_{0,6})\,,\\
&\wp:\CFB^{\rm reg}(\CV_{[6]}',C_{0,{6}})\rightarrow \CFB^{\rm reg}(\CV_{[4]},C_{0,4})\,,
\end{aligned}
\end{equation}
in a natural way. The Verlinde loop operators can then be defined as the composition
\begin{equation}
\SV_\ga:=\wp\circ\SM_\ga\circ\imath\,.
\end{equation}
Concretely this boils down to taking a certain linear combination of the matrix elements
$\SM_\ga(\be,\ST_\be)_{ss'}^{s''}$ representing the monodromy along $\ga$ on
$\CFB^{\rm reg}(\CV_{[6]},C_{0,6})$.
The explicit calculations of the operator $\SV_{\ga}$ in \cite{AGGTV,DGOT}
shows that the Verlinde loop operators define a representation of 
 ${\rm Fun}_q(\CM_{\rm flat}(C))$ on the space of conformal blocks 
 $\CFB^{\rm reg}(\CV_{[4]},C_{0,n})$ which is equivalent to the one defined in
 Section \ref{reprs} if we identify variables
 \begin{equation}
 \be\,=\,\frac{Q}{2}+{\mathrm i}\frac{l}{4\pi b}\,,\qquad
 \al_k\,=\,\frac{Q}{2}+{\mathrm i}\frac{l_k}{4\pi b}\,,\quad k=1,\dots,4.
 \end{equation}
Let us summarize the observations made in this section in the following table:
\begin{center}
\begin{tabular}{l|l}
Quantisation of ... & ... is realised in CFT via \\ \hline\hline \\[-2ex]
Darboux coordinates $(q,h)$ & conformal Ward indentities \\[1ex]
Fenchel-Nielsen coordinates $(k,l)$ & Verlinde loop operators 
\\[1ex] \hline 
\end{tabular}
\end{center}
{\it The degenerate fields have a beautiful interpretation in the 
gauge theories $\CG_C$ in terms of a family of observables called
surface operators \cite{AGGTV}, explained also 
in the article \cite{G} in this collection.}

\subsection{Liouville conformal blocks as solutions to the Riemann-Hilbert problem}\label{sec:RHsol}

We claim that the solution to the Riemann-Hilbert type problem defined
in Section \ref{loop-S-dual} is given by the Liouville conformal blocks in 
the following sense 
\begin{equation}\label{AGT-fin}
\CZ^{\rm inst}_\si(a,m;\tau;\ep_1,\ep_2)\,=\,
\CZ^{\rm spur}_\si(\al;\tau;b)\,\CZ^{\rm\sst Liou}_{\si}(\be,\al;q;b)\,,
\end{equation}
The solution of the Riemann-Hilbert problem defined
in Section \ref{loop-S-dual} is unique up to 
multiplication with meromorphic functions which may have
poles only at the
boundary of $\CM(C_{0,4})$. The resulting freedom can be absorbed into 
$\CZ^{\rm spur}_\si(\al,\tau;b)$.

In order to verify \rf{AGT-fin} we need to show that 
the representation of the mapping class group on spaces of 
Liouville conformal blocks is the same as the one coming from the
quantum theory of $\CM_{\rm flat}^{\BR,0}(C)$ as described in Section 
\ref{q-Mflat}. This boils down to the comparison of the respective
realizations of $B$ and $F$-moves. The coincidence of $B$-moves
is trivial to verify. The realization
of the F-move on Liouville conformal blocks 
was calculated in \cite{T01}, where a relation of the form
\begin{equation}\label{fustrsf}
\CZ_{s}^{\rm\sst Liou}(\be_1,q) \,=\,
\int_\BS d\be_2\;\Fus{\al_1}{\al_2}{\al_3}{\al_4}{\be_{1}}{\be_2}\,
\CZ_{t}^{\rm\sst Liou}(\be_2,q)\,,\quad\BS\equiv \frac{Q}{2}+i\BR^+\,,
\end{equation}
was found. For the normalization defined in \rf{Ndef} we find the {\it same}
kernel $\Fus{\al_1}{\al_2}{\al_3}{\al_4}{\be_{1}}{\be_2}$ in the
relation \rf{fustrsf}, as was found within the
quantum theory of $\CM_{\rm flat}^{\BR,0}(C)$ described in Section 
\ref{q-Mflat}.

This is good enough to conclude that \rf{AGT-fin} must hold. 
To round off the picture, let us exhibit the meaning of the partition 
functions $\CZ^{\rm\sst Liou}_{\si}(\be,\al;q;b)$ within the quantisation of 
$\CM_{\rm flat}^{\BR,0}(C)$.

In this section we have described two different representations for the 
quantisation of one
and the same Poisson-manifold, obtained by quantisation of the coordinates
$(q,h)$ and $(l,k)$, respectively.
One may expect that these two representations should be unitarily 
equivalent. 
The eigenstates $|l\rangle$ of the operator $\sll$ 
are complete in $\CH_{\si_s}$ . It should therefore be possible
to relate the wave-function  $\psi(q)\equiv 
\langle q|\psi\rangle$ representing a state $|\psi\rangle$ 
in the holomorphic representation to the wave-function 
$\Psi(l)\equiv \langle l|\psi\rangle$ representing the same state 
in the length representation as
\begin{equation}
\psi(q)\equiv \langle\, q\,|\,\psi\,\rangle
\,=\,\int dl\;\langle\, q\,|\,l\,\rangle \langle\, l\,|\,\psi\,\rangle\,.
\end{equation}
The kernel $\langle\, q\,|\,l\,\rangle$ is the complex conjugate of
the wave-function $\langle\, l\,|\,q\,\rangle$
of the ``coherent'' state $|\,q\,\rangle$ in the length 
representation.

%It follows from the discussion above that one
%may always interpret the holomorphic wave-functions $\psi(q)$
%representing states in the coherent state representation for 
%the quantum theory of $\CM_{\rm flat}^0(C)$ as  chiral 
%partition functions $\CZ^{\rm Liou}(\CF,C)$ of some conformal block $\CF$.
%This holds in particular for the wave-functions $\langle\, q\,|\,l\,\rangle$
%of the states $|\,l\,\rangle$.

One may use essentially the  
same arguments as presented in Section \ref{loop-S-dual} 
to conclude that $\langle\, q\,|\,l\,\rangle$ must solve the same
Riemann-Hilbert problem as discussed above. Combined with a discussion of the 
asymptotics at the boundary of $\CT(C)$ we may thereby conclude \cite{T,TV2} 
that 
\begin{equation}
\langle\, q\,|\,l\,\rangle\,=\,\CZ^{\rm\sst Liou}_{\si}(\be,\al;q;b)\,.
\end{equation}
We have thereby identified more
precisely which wave-functions the conformal blocks are:
They describe the change of representation between the two 
natural representations for the quantum theory
of $\CM_{\rm flat}^{\BR,0}(C)$ discussed in this section.

\subsection{The Nekrasov-Shatashvili limit}

The results reviewed in this article are related to the work \cite{NRS} in an interesting way.
In order to explain the relations to \cite{NRS} let us consider the limit $\ep_2\ra 0$ corresponding to the classical
limit for the quantum theory discussed in the previous sections. This will also provide
further insight into the meaning of $\langle\, q\,|\,l\,\rangle=\CZ^{\rm\sst Liou}$.
It can be shown \cite{T10} that the conformal blocks
behave as 
\begin{equation}\label{NSlim}
\log\CZ^{\rm\sst Liou}_{\si}(\be,\al;q;b)
\sim -\frac{1}{\ep_2}
\CY(l,m;q;\ep_1)\,,
\end{equation}
assuming that the variables are related by \rf{paramid}.
$\CY(l,m;q;\ep_1)$ is defined as follows: For given values of $l$ and $q$ let
us consider differential operators of the form $\ep_1^2(\pa_y^2+t(y))$, 
with $t(y)$ of the form \rf{fouroper}. It can be argued that there is a unique 
choice $h=h(l,q)$ 
for the residue $h$ such that $\ep_1^2(\pa_y^2+t(y))$ has monodromy 
with trace equal to $2\cosh(l/2)$. $\CY(l,m;q;\ep_1)$ is defined up to 
a constant by the condition that 
\begin{equation}\label{hfromY}
\frac{\pa}{\pa q}\CY(l,m;q;\ep_1)\,=\,-h(l,q)\,.
\end{equation}
The constant can be fixed by demanding that the constant term $\CY_0(l,m;\ep_1)$ in
\begin{equation}
\CY(l,m;q;\ep_1)\underset{q\ra 0}{\sim} -(\de-\de_1-\de_2)\log q+
\CY_0(l,m;\ep_1)+\CO(q)\,,
\end{equation}
is one half of the sum of the Liouville actions on the three-punctured spheres
into which $C_{0,4}$ splits when $q\ra 0$.
We furthermore have 
\begin{equation}\label{kfromY}
k(l,q)\,=\,4\pi i\,\frac{\pa}{\pa l}\,\CY(l,m;q;\ep_1)\,.
\end{equation}
To verify \rf{kfromY} note, on the one hand,
that the Verlinde loop operators reduce to the trace functions
in the limit $\ep_2\ra 0$. 
Recall that the trace functions  may be parameterised 
by the (complexified) Fenchel-Nielsen coordinates $(l,k)$. 
The resulting expression
may be compared to one following from \rf{rep0,4} and \rf{NSlim} in this limit,
giving \rf{kfromY}.

The pairs of coordinates $(l,k(l,q))$ describe a Lagrangian sub-manifold denoted 
${\rm Op}_{\fsl_2}(C)$
within $\CM_{\rm flat}(C)$ sometimes called the "brane of opers". 
It follows
from \rf{kfromY}, \rf{hfromY} that $\CY(l,m;q;\ep_1)$ is the generating
function of this sub-manifold.
We thereby arrive at the description for the $\ep_2$-limit of the instanton
partition functions that was proposed in \cite{NRS}. 
%It is here found to be
%a consequence of the fact that the Liouville conformal blocks
%represent the solutions to the Riemann-Hilbert problem characterising 
%the instanton partition functions. 
One may therefore view the results of \cite{TV2} reviewed in this
article as the generalisation of the results from \cite{NRS}
to nonzero $\ep_2$.

\subsection{Quantization of Seiberg-Witten theory}

It will furthermore be instructive to consider the limit where
both $\ep_1,\ep_2\ra 0$, in which ${\BBE}_{\ep_1,\ep_2}^4$ turns
into $\BR^4$, and we can make contact with 
Seiberg Witten theory.

To begin with, let us note that
$\ep_1\pa_y^2+t(y)$ turns into the
quadratic differential $-\vt(y)$ when $\ep_1\ra 0$.
Using $\vt(y)$ we define the Seiberg-Witten curve $\Sigma$ 
as usual by
\begin{equation}
\Sigma\,=\,\{\,(v,u)\,|\,v^2=\vt(u)\,\}\,.
\end{equation}
It follows by WKB analysis of the differential equation 
$(\ep_1\pa_y^2+t(y))\chi=0$ that
the coordinates $l_e$ have asymptotics that can be expressed in terms of
the Seiberg-Witten differential  $\Lambda$ on $\Sigma$ 
defined such that $\Lambda^2=\vt(u) (du)^2$.
We find
\begin{equation}
l\,\sim\,\frac{4\pi}{{\ep_\1}}\,a\,,\qquad
k\,\sim\,\frac{4\pi}{{{\ep_\1}}}\,a^{\rm\sst D}\,,
\end{equation}
where $a$ and $a^{\rm\sst D}$ are periods of the 
Seiberg-Witten differential  $\Lambda$ defined as
\begin{equation}
 a:=\int_{\al}\La\,, \qquad a^{\rm\sst D}:
=\int_{\be}\La\,,
\end{equation}
with $\al$ and $\be$ 
being lifts of  $\ga_s$ and $\ga_t$ to 
cycles on $\Sigma$ that project to zero,
respectively.

The prepotential $\CF$ is obtained in the limit 
$\ep_1,\ep_2\ra 0$ as follows:
\begin{equation}\begin{aligned}
\CF(a,m,q):& =\lim_{\ep_1,\ep_2\ra 0}\ep_1\ep_2
\CZ^{\rm inst}_\si(a,m,q;\ep_1,\ep_2)\\
& = \lim_{\ep_1,\ep_2\ra 0}\ep_1\CY(a,m;q;\ep_1)\,.
\end{aligned} \end{equation}
$\CF(a,m;q)$  satisfies the relations
\begin{equation}\label{Matone}
a^{\rm D}\,=\,\frac{\pa}{\pa a}\CF(a,q)\,,
\qquad h\,=\,-\frac{\pa}{\pa q}\CF(a,q)
\,.
\end{equation} 
A proof of the relations \rf{Matone} that is valid 
for all $A_1$-theories of class $\CS$
was given in \cite[Section 7.3.2]{GT}. The relations \rf{Matone} are 
equivalent to the statement
that both the coordinates $(a,a^D)$ describing the 
special geometry underlying Seiberg-Witten theory, and
the coordinates $(q,h)$ introduced above can be
seen as systems of Darboux coordinates for the same
space $T^*\CT(C)$. The prepotential $\CF(a,m,q)$ is nothing but 
the generating function of the change of variables
between $(a,a^D)$ and $(q,h)$. 

These observations show that the relations
between the quantum theory on $\CM^0_{\rm flat}(C)$ 
and the gauge theories $\CG_C$ 
discussed in this article can be seen as the
quantization of the special geometry used in  
Seiberg-Witten theory. The dual zero modes
$a$ and $a^{\rm D}$ turn into the Darboux coordinates $l$ and $k$ 
upon partial compactification to $S^1\times \BR^3$ or
${\BBE}_{\ep_1}^2\times \BR^2$.
Further compactification to a four-ellipsoid leads to the 
quantization of these
zero modes.  

{\it
It is intriguing to observe that very similar 
ideas have been discussed in the context of topological string
theory, where it has been proposed that the partition 
function of the topological string has an interpretation as a
wave-function arising from the quantization of special 
geometry. The geometric engineering of gauge theories 
within string theory leads to relations between topological 
string and instanton partition functions, see the
articles  \cite{A,KW} in this volume for a review. 
One may hope that
the relations with the quantization of moduli spaces of
vacua discussed in this article may help us to get a more
unified picture.}

{\bf Acknowledgements} The author would like to thank T. Dimofte, M. Gabella, A. Neitzke 
and T. Okuda for very useful remarks on a preliminary version of this article.

%\newpage
\appendix
\section{Riemann surfaces: Basic definitions and results}
\label{RS}
\setcounter{equation}{0}

This appendix introduces 
basic definitions and results
concerning Riemann surfaces $C$ that will be used throughout the paper.
A Riemann surface $C$ is a two-dimensional topological surface $S$ together
with a choice of complex structure on $S$.
We will denote by 
$\CM(S)$ the moduli space of complex structures
on a two-dimensional surface $S$, and by 
$\CT(C)$ the Teichm\"uller space of deformations of complex structures on 
the Riemann surface $C$.

%We begin by introducing some basic definitions
%concerning Riemann surfaces that will be used throughout the paper.
%We will mostly 
%consider Riemann surfaces $C\sim C_{0,n}$ of genus zero with 
%$n$ marked points. 
%Such Riemann surfaces can always be represented as 
%$C\simeq \BP^1\setminus\{z_1,\dots,z_n\}$. 
%We will furthermore represent $\BP^1$ as $\BC\cup \{\infty\}$.
%The standard coordinate $y$ on $\BC$ will be used as 
%our preferred coordinate for $C$. 
%Surfaces $C$ that are related by Moebius transformations
%\[
%y\,\mapsto\,\frac{ay+b}{cy+d}\,,
%\]
%will be identified. Without loss of generality 
%we may therefore assume that $z_n=\infty$, $z_{n-1}=1$, $z_1=0$,
%and use $(z_{n-2},\dots,z_2)$ as coordinates for the 
%moduli space $\CM_{n}$ of complex structures on $C$.

\subsection{Complex analytic gluing construction}
\label{sec:glueing}

A convenient family of particular coordinates on the Teichm\"uller spaces
$\CT(C)$ is defined by means of the complex-analytic 
gluing construction of Riemann surfaces $C$ from three punctured
spheres \cite{Ma,HV}. Let us  briefly review this construction.

Let $C$ be a (possibly disconnected) Riemann surface. 
Fix a complex number $q$ with $|q|<1$, and
pick two points $Q_1$ and $Q_2$ on $C$
together with coordinates $z_i(P)$ in a neighborhood of 
$Q_i$, $i=1,2$, such that $z_i(Q_i)=0$, and such that the discs $D_i$,
\[
D_i\,:=\,\{\,P_i\in C\,;\,|z_i(P_i)|<|q|^{-\frac{1}{2}}\,\}\,,\qquad i=1,2\,,
\]
do not intersect.
One may define the annuli $A_i$,
\[
A_i\,:=\,\{\,P_i\in C\,;\,|q|^{\frac{1}{2}}<|z_i(P_i)|<|q|^{-\frac{1}{2}}\,\}\,,\qquad i=1,2\,.
\]
To glue $A_1$ to $A_2$ let us identify two points $P_1$ and $P_2$ on $A_1$ and $A_2$, 
respectively, iff the coordinates of these two points satisfy the equation
\begin{equation}\label{glueid}
z_1(P_1)z_2(P_2)\,=\,q\,.
\end{equation}
If $C$ is connected one creates an additional handle, and if $C=C_1\sqcup C_2$
has two connected components one gets a single connected component after 
performing the gluing operation. In the limiting case where $q=0$ 
one gets a nodal surface which represents a component of 
the boundary $\pa\CM(S)$ defined by the Deligne-Mumford 
compactification $\overline{\CM}(S)$.

By iterating the gluing operation one may build any 
Riemann surface $C$ of genus $g$ with $n$ punctures 
from three-punctured spheres $C_{0,3}$. 
Embedded into $C$ we naturally get a collection of annuli 
$A_1,\dots,A_h$, where
\begin{equation}
h\,:=\,3g-3+n\,.
\end{equation}
The construction above can be used to define 
a $3g-3+n$-parametric family of Riemann
surfaces, parameterised by a collection $q=(q_1,\dots,q_h)$ 
of complex parameters. These parameters can be taken
as coordinates for a neighbourhood of 
a component in the boundary $\pa\overline{\CM}(S)$ which are 
complex-analytic with respect to 
its natural complex structure \cite{Ma}. 

Conversely, assume given a Riemann surface $C$ and a 
cut system, a collection 
$\CC=\{\ga_1,\dots,\ga_h\}$ of 
homotopy classes of non-intersecting simple closed curves
on $C$.  
Cutting along all the curves in $\CC$ produces a pants decompostion,
$C\setminus\CC\simeq\bigsqcup_{v}C_{0,3}^v$, where the $C_{0,3}^v$
are three-holed spheres.

Having glued $C$ from three-punctured spheres defines a
distinguished cut system, defined by a collection
of simple closed curves $\CC=\{\ga_1,\dots,\ga_h\}$ 
such that $\ga_r$ can be embedded into the annulus $A_r$
for $r=1,\dots,h$. 

\begin{figure}[htb]
\epsfxsize3cm
\centerline{\epsfbox{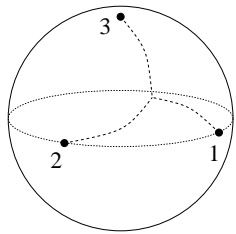}}
\caption{\it A sphere with three punctures, and 
a choice of branch cuts for the definition of 
the logarithms of local coordinates around the punctures.}
\label{threept}\vspace{.3cm}
\end{figure}

An important deformation of the complex structure of $C$ is 
the Dehn-twist: It corresponds to rotating one end of an 
annulus $A_r$ by $2\pi$ before regluing, and can be described by
a change of the local coordinates used in the gluing
construction. The coordinate $q_r$ can not distinguish complex
structures related by a Dehn twist in $A_r$. It is often
useful to replace the coordinates $q_r$
by logarithmic coordinates $\tau_r$ such that $q_r=e^{2\pi i \tau_r}$. 
This corresponds to replacing the gluing identification \rf{glueid} 
by its logarithm. In order to define the logarithms of the coordinates
$z_i$ used in \rf{glueid}, one needs to introduce branch cuts
on the three-punctured spheres, an example being
depicted in Figure \ref{threept}.

By imposing the requirement that the branch cuts chosen on each
three-punctured sphere glue to a connected three-valent 
graph $\Ga$ on $C$, one gets an unambiguous definition of the
coordinates $\tau_r$. We see that the logarithmic versions of
the gluing construction that define the coordinates $\tau_r$
are parameterized by the pair of data $\si=(\CC_\si,\Ga_\si)$,
where $\CC_\si$ is the cut system defined by the gluing construction,
and $\Ga_\si$ is the three-valent graph specifying the choices 
of branch cuts. In order to have a handy terminology we will 
call the pair of data  $\si=(\CC_\si,\Ga_\si)$ a {\it pants decomposition},
and the three-valent graph $\Ga_\si$ will be called the Moore-Seiberg graph, 
or MS-graph associated to a pants decomposition $\si$. The construction 
outlined above gives a set of coordinates for the neighbourhood $\CU_\si$ of the boundary component of
$\CT(C)$ corresponding to $\si$.

The gluing construction depends on the choices of 
coordinates around the punctures
$Q_i$.  There exists an ample supply of choices
for the coordinates $z_i$ such that the union of the 
neighbourhoods $\CU_{\si}$ produces a cover of $\CM(C)$ \cite{HV}. 
For a fixed choice of these coordinates
one produces families of Riemann surfaces fibred over
the multi-discs $\CU_{\si}$ with coordinates $q$.
Changing the coordinates $z_i$ around $Q_i$ produces a 
family of Riemann surfaces which
is locally biholomorphic to the initial one \cite{RS}. 

\subsection{The Moore-Seiberg groupoid}\label{MSdef}

Let us note \cite{MS,BK} that
any two 
different pants decompositions 
$\si_2$, $\si_1$ can be connected by a sequence of
elementary moves localized in subsurfaces of $C_{g,n}$ of type
$C_{0,3}$, $C_{0,4}$ and $C_{1,1}$. The elementary moves are called the 
$B$, $F$, $Z$ and $S$-moves, respectively.
Graphical representations for the elementary 
moves $F$, $S$ and $B$  are given in 
Figures \ref{fmove}, \ref{smove}  and \ref{bmove}, respectively. 
The $Z$-move is just the change of
distinguished boundary component in a three-punctured sphere.

One may formalize
the resulting structure by introducing a two-dimensional 
CW complex $\CM(C)$ with set of vertices $\CM_\0(C)$
given by the
pants decompositions $\si$, and a 
set of edges $\CM_\1(C)$ associated to the elementary
moves. 
The Moore-Seiberg groupoid is defined to be the path groupoid of 
 $\CM(C)$. It can be described in terms of generators and relations,
the generators being associated with the edges of  $\CM(C)$, 
and the relations associated with the faces of $\CM(C)$. 
The classification of the relations was first presented in \cite{MS},
and rigorous mathematical proofs have been presented in \cite{FG,BK}.
The relations are all represented by sequences of elementary moves localized
in subsurfaces $C_{g,n}$ with genus $g=0$ and $n=3,4,5$ punctures,
as well as $g=1$, $n=1,2$. Graphical representations of the
relations can be found in \cite{MS,FG,BK}.

\subsection{Uniformization}

The classical uniformization theorem ensures existence and uniqueness of 
a hyperbolic metric, 
a metric of constant negative curvature, on a Riemann surface $C$.
In a local chart with complex analytic coordinates $y$
one may represent this metric in the form $ds^2=e^{2\vf}dyd\bar y$, with
$\vf$ being a solution to the Liouville equation
$\pa\bar \pa\vf=\mu e^{2\vf}dy d\bar y$.

The solutions to the Liouville equation
may be parameterized by a function $t(y)$ related to $\vf$ as
\begin{equation}\label{tfromphi}
t:=\,-(\pa_y\vf)^2+\pa_y^2\vf\,.
\end{equation}
$t(y)$ is holomorphic as a consequence of the Liouville equation.
%\begin{equation}\label{opertrsf-a}
%t(y)\;\mapsto\;(y'(w))^2t(y(w))+\frac{1}{2}\{y,w\}\,,
%\end{equation}
%where the 
%Schwarzian derivative $\{y,w\}$ is defined as
%\begin{equation}
%\{y,w\}\,\equiv\,\left(\frac{y''}{y'}\right)'-
%\frac{1}{2}\left(\frac{y''}{y'}\right)^2\,.
%\end{equation}
%Equation \rf{opertrsf-a}  
%is the transformation law characteristic for {\it projective}
%connections, which are also called $\fsl_2$-opers, or opers for short.
The solution to the Liouville equation can be reconstructed from 
$t(y)$ by first finding 
the solutions to 
\begin{equation}\label{DFuchsian}
(\pa_y^2+t(y))\chi=0\,.
\end{equation}
Picking two linearly independent solutions $\chi_\pm$ of \rf{DFuchsian} with
$\chi_+'\chi_--\chi_-'\chi_+=1$ allows us to represent
$e^{2\vf}$ as $e^{2\vf}=-(\chi_+\bar\chi_--\chi_-\bar\chi_+)^{-2}$.
The  hyperbolic metric $ds^2=e^{2\vf}dyd\bar y$ may then be written
in terms of the quotient $A(y):=\chi_+/\chi_-$ as
\begin{equation}
ds^2\,=\,e^{2\vf}dyd\bar y\,=\,\frac{\pa A\bar\pa {\bar A}}{({\rm Im}(A))^2}\,.
\end{equation}
It follows that $A(y)$ represents a conformal mapping from $C$ to 
a domain $\Omega$ 
in the upper half plane $\BU$ with its standard constant curvature 
metric. The monodromies of the solution $\chi$ are represented on $A(y)$
by Moebius transformations. These Moebius transformations describe the 
identifications of the boundaries of the simply-connected domain $\Omega$ in $\BU$
which represents the image of $C$ under $A$. 
$C$ is therefore conformal to $\BU/\Ga$, where
the Fuchsian group $\Ga$ is the monodromy group of the 
differential operator $\pa_y^2+t(y)$.

\section{Moduli spaces of flat connections}\label{Mflat}

\setcounter{equation}{0}

 In this appendix we shall review some of the basic definitions and
results concerning the moduli spaces $\CM_{\rm flat}(C)$.

\subsection{Moduli of flat connections and character variety}\label{sec:loops}

We will consider flat ${\rm PSL}(2,\BC)$-connections $\nabla=d-A$
on Riemann surfaces $C$. Let $\CM_{\rm flat}(C)$ be the moduli space
of all such connections modulo gauge transformations.

Given a flat ${\rm PSL}(2,\BC)$-connection $\nabla=d-A$, one may define 
its holonomy $\rho(\ga)$ along a closed loop $\ga$ 
as $\rho(\ga)=\CP\exp(\int_\ga A)$. The assignment
$\ga\mapsto \rho(\ga)$ defines a representation of $\pi_1(C)$ in 
${\rm PSL}(2,\BC)$.
As any flat connection is locally gauge-equivalent to the trivial connection,
one may characterize gauge-equivalence classes of flat connections
by the corresponding representations $\rho:\pi_1(C)\ra {\rm PSL}(2,\BC)$.
This allows us to 
identify the moduli space $\CM_{\rm flat}(C)$ of
flat ${\rm PSL}(2,\BC)$-connections on $C$
with the so-called
character variety 
\begin{equation}
\CM_{\rm char}(C):={\rm Hom}(\pi_1(C),{\rm PSL}(2,\BC))/{\rm PSL}(2,\BC)\,.
\end{equation}
The moduli space $\CM_{\rm flat}(C)$ has a natural real slice, the moduli space $\homslr$ of flat 
${\rm PSL}(2,\BR)$-connections. 

\subsection{The Teichm\"uller component}\label{uniflat}

There is a well-known relation between the Teichm\"uller space $\CT(C)$
and a connected component of the moduli space 
$\CM_{\rm flat}^{\BR}(C)$ of flat ${\rm PSL}(2,\BR)$-connections 
on $C$. This component is called the Teichm\"uller component and will be denoted 
as $\CM_{\rm flat}^{\BR,0}(C)$. The relation between $\CT(C)$  and
$\CM_{\rm flat}^0(C)$ may be described as follows.
To a hyperbolic metric  $ds^2=e^{2\vf}dyd\bar y$ let us 
associate the connection $\nabla=\nabla'+\nabla''$, 
\begin{equation}\label{hypconn}
\nabla''=\bar\pa\,,\qquad\nabla'\,=\,\pa+M(y)dy\,, \qquad
M(y)\,=\,\bigg(\,\begin{matrix} 0 & -t \\ 1 & 0 \end{matrix}\,\bigg)\,,
\end{equation}
with $t$ constructed from $\vf(y,\bar y)$ as in \rf{tfromphi}.
This connection is flat since $\pa_y\bar{\pa}_{\bar y}\vf
=\mu e^{2\vf}$
implies $\bar\pa t=0$. The Fuchsian group $\Ga$ characterizing the
uniformization of $C$ is nothing but the
holonomy $\rho$ of the connection  $\nabla$ defined in \rf{hypconn}.
%The form \rf{hypconn} of $\nabla$
%is preserved by changes
%of local coordinates if $t=t(y)$ transforms via \rf{opertrsf-a}.

The Fuchsian groups $\Ga$ fill out the
connected component 
$\CM_{\rm char}^{\BR,0}(C)\simeq \CT(C)$ 
in $\homslr$ called the 
Teichm\"uller component.

\subsection{Fock--Goncharov coordinates}\label{FGcoords}

Let $\tau$ be a triangulation of the surface $C$ such that
all vertices coincide with marked points on $C$. An edge
$e$ of $\tau$ separates two triangles defining a
quadrilateral $Q_e$ with corners being the marked points
$P_1,\ldots,P_4$. For a given local system $(\CE,\nabla)$,
let us choose four sections $s_i$, $i=1,2,3,4$ that 
%are holomorphic in $Q_e$, 
obey the condition
%\begin{equation}
$\nabla s_i % = \left(\frac{\pa}{\pa y}-A(y)\right)s_i 
= 0,$
%\end{equation}
and are eigenvectors of the monodromy around $P_i$. Out of
the sections $s_i$ form \cite{FG,GMN2}
\begin{align}
\CX_e^{\tau} := -\frac{(s_1\wedge s_2)(s_3\wedge s_4)}{(s_2\wedge s_3)(s_4\wedge s_1)},
\end{align}
where all sections are evaluated at a common point $P\in
Q_e$. It is not hard to see that $\CX_e^{\tau}$ does not
depend on the choice of $P$.

There exists a simple description of  
the relations between  the coordinates
associated to different triangulations. If  
triangulation $\tau_e$ 
is obtained from $\tau$ by changing only the diagonal in the quadrangle
containing $e$, we have 
\begin{equation}\label{cluster}
\CX_{e'}^{\tau_e}\,=\,\left\{
\begin{aligned} &\CX_{e'}^{\tau}\,
\big(1+
(\CX_e^{\tau})^{-{\rm sgn}(n_{e'e})}\big)^{-n_{e'e}} \;\;&{\rm if}\;\;e'\neq e\,,\\
&(\CX_e^{\tau})^{-1}\;\;&{\rm if}\;\;{e'= e}\,.
\end{aligned}\right.
\end{equation}
This reflects part of the structure of 
a cluster algebra that $\CM_{\rm flat}^{}(C)$ 
has.

\subsection{Trace functions}

The trace functions
\begin{equation}\label{tracedef}
L_\ga:=\nu_\ga{\rm tr}(\rho(\ga))\,,
\end{equation}
represent useful coordinate functions for $\homsl$. The signs 
$\nu_\ga$ will be chosen such that the restriction to $L_{\ga}$ to the
Teichm\"uller component $\CM_{\rm char}^{\BR,0}(C)$ satisfies 
$L_{\ga}=2\cosh(l_\ga/2)>2$, where $l_\ga$ is the length of the hyperbolic geodesic
on $\BU/\Ga$ isotopic to $\ga$.

% \subsection{Skein algebra}
The coordinate functions $L_\ga$ generate the 
commutative algebra $\CA(C)\simeq {\rm Fun}^{\rm alg}(\CM_{\rm flat}(C))$ 
of functions  on $\CM_{\rm flat}(C)$. 
The well-known relation ${\rm tr}(g){\rm tr}(h)={\rm tr}(gh)+{\rm tr}(gh^{-1})$
valid for any pair of $SL(2)$-matrices $g,h$ implies that the
geodesic length functions satisfy the 
 so-called skein relations,
\begin{equation}\label{skeinrel}
L_{\ga_1} L_{\ga_2}\,=\,L_{S(\ga_1,\ga_2)}\,,
\end{equation}
where $S(\ga_1,\ga_2)$ is the loop obtained from 
$\ga_1$, $\ga_2$ by means of the smoothing operation,
defined as follows. The application of $S$ to a single intersection
point of $\ga_1$, $\ga_2$ is depicted in 
Figure \ref{skeinfig} below.
\begin{figure}[htb]
\epsfxsize8cm
\centerline{\epsfbox{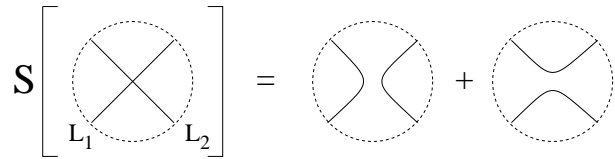}}
\caption{The symmetric 
smoothing operation}\label{skeinfig}\vspace{.3cm}
\end{figure}
The general result is obtained by
applying this rule at each intersection point, and summing the
results.

\subsection{Topological classification of closed loops}\label{Dehn}

With the help of pants decompositions one may conveniently classify all
non-selfintersecting closed loops on $C$ up to homotopy. To a loop $\ga$
let us associate the collection of integers $(r_e,s_e)$ 
associated to all edges $e$ of $\Ga_{\si}$ which are defined as follows.
Recall that there is a unique curve $\ga_e\in\CC_\si$ that intersects 
a given edge $e$ on $\Ga_\si$ exactly once, and which does not intersect
any other edge. The integer $r_e$ is defined as 
the number of intersections between $\ga$ and 
the curve $\ga_e$. Having chosen an orientation 
for the edge $e_r$ we will define
$s_e$ to be the intersection index between $e$ and $\ga$. 

Dehn's theorem (see \cite{DMO} for a nice discussion)
ensures that the curve $\ga$ is up to homotopy uniquely classified by the
collection of integers $(r,s)$, subject to the restrictions 
\begin{equation}
\begin{aligned}
{\rm (i)} \quad & 
r_e\geq 0\,,\\ {\rm (ii)} \quad & {\rm if}\;\;r_e=0\;\Rightarrow\;s_e\geq 0\,,\\
{\rm (iii)} \quad &
r_{e_\1}+r_{e_\2}+r_{e_\3}\in 2\BZ\;\,{\rm whenever}\;\,
\ga_{e_\1},\ga_{e_\2},\ga_{e_\3}\;\,\text{bound the same trinion}.
\end{aligned}
\end{equation}
We will use the notation $\ga_{(r,s)}$ for the geodesic which has
parameters $(r,s):e\mapsto (r_e,s_e)$.

\subsection{Generators and relations} \label{sec:Genrel}

The pants decompositions allow us to describe $\CA(C)$ 
in terms of generators and relations. 
As set of generators for $\CA(C)$ 
one may take the functions
$L_{(r,s)}\equiv L_{\ga_{(r,s)}}$. 
The skein relations imply various relations among the 
$L_{(r,s)}$. It is not hard to see that these relations
allow one to express arbitrary $L_{(r,s)}$ in terms
of a finite subset of the set of $L_{(r,s)}$. 

Let us temporarily restrict attention to surfaces with genus zero and 
$n=4$ boundaries. 
The Moore-Seiberg graph $\Ga_{\si}$ will then have only
one internal edge, allowing us to
drop the index $e$ labelling the edges.   
Let us introduce the geodesics $\ga_s=\ga_{(1,0)}$, $\ga_t=\ga_{(0,2)}$
and $\ga_u=\ga_{(1,2)}$.
The geodesics $\ga_s$ and $\ga_t$
are depicted as red curves on the left and 
right half of Figure \ref{fmove}. 
We will denote
$L_{k}\equiv L_{\ga_k}$, where $k\in\{s,t,u\}$.
The trace functions $L_s$, $L_t$ and $L_u$ 
generate $\CA(C)$.

These coordinates are not independent, though. Further relations
follow from the relations in $\pi_1(C)$.
It can be shown (see e.g. \cite{Go09} for a review) 
that the
coordinate functions $L_s$, $L_t$ and $L_u$
satisfy an algebraic
relation of the form 
\begin{subequations}\label{algrel}
\begin{equation}
P(L_s,L_t,L_u)\,=\,0\,.
\end{equation}
The polynomial $P$ in \rf{algrel} is
explicitly given
as\footnote{Comparing to \cite{Go09} note
that some signs were absorbed by a suitable choice of the signs
$\nu_\ga$ in \rf{tracedef}.} 
\begin{align}
 P(L_s, L_t, L_u) := &-L_s L_t L_u + L_s^2 + L_t^2 + L_u^2 \nonumber \\
& +L_s (L_3L_4 + L_1L_2) + L_t (L_2L_3 + L_1L_4) + L_u (L_1L_3 + L_2L_4)
\nonumber \\ &
-4 + L_1^2+L_2^2+L_3^2+L_4^2+L_1L_2L_3L_4\,.
\end{align}
\end{subequations}
In the expressions above we have denoted $L_{i}:=L_{\ga_i}$, where
$\ga_i$, $i=1,2,3,4$ represent 
the boundary components of $C_{0,4}$, labelled according
to the convention defined in Figure \ref{fmove}.

\subsection{Trace functions in terms of Fock-Goncharov coordinates}\label{trace-FG}

Assume given a path $\varpi_{\ga}$ on the fat graph
homotopic to a simple closed curve $\ga$ on $C_{g,n}$. Let
the edges be labelled $e_i$, $i=1,\ldots,r$ according to
the order in which they appear on $\varpi_{\ga}$, and
define $\si_i$ to be $1$ if the path turns left at the
vertex that connects edges $e_i$ and $e_{i+1}$, and to be
equal to $-1$ otherwise. Consider the following matrix,
\begin{equation}\label{fuchsgen}
{\rm X}_{\ga} = {\rm V}^{\si_r}{\rm E}(z_{e_r})\cdots {\rm V}^{\si_1}
{\rm E}(z_{e_1}),
\end{equation}
where $z_e=\log X_e$, and the matrices ${\rm E}(z)$ and
${\rm V}$ are defined respectively by
\begin{equation}
{\rm E}(z) = \bigg(\begin{array}{cc} 0 & +e^{+\frac{z}{2}}\\
-e^{-\frac{z}{2}} & 0 \end{array}\bigg),\quad
{\rm V} = \bigg(\begin{array}{cc} 1 & 1 \\ -1 & 0 \end{array}\bigg).
\end{equation}
Taking the trace of ${\rm X}_{\ga}$ one gets the hyperbolic
length of the closed geodesic isotopic to $\ga$
via~\cite{F97}
\begin{equation}\label{glength}
L_{\ga} \equiv 2\cosh (l_{\ga}/2) = |{\rm tr}({\rm X}_{\ga})|.
\end{equation}

We may observe that the classical expression for
$L_{\ga}\equiv 2\cosh\frac{1}{2}l_{\ga}$ as given by
formula \ref{glength} is a linear combination of monomials
in the variables $u_e^{\pm1}\equiv e^{\pm\frac{z_e}{2}}$ of
the very particular form \rf{L-X}.% ,\vspace*{-5pt}
%\begin{equation}\label{Lclass}
%L_{\ga} = \sum_{\nu\in\CF} C_{\tau,\ga}(\nu) \prod_e u_e^{\nu_e}\vspace*{-5pt}
%\end{equation}
%The
%coefficients $C_{\tau,\ga}(\nu)$ are positive integers.

\subsection{Fenchel-Nielsen coordinates for 
$\CM_{\rm flat}^{\BR,0}(C)$} \label{sec:FN}
\label{sec:2hyp}

One may express $L_s$, $L_t$ and $L_u$
in terms of the Fenchel-Nielsen coordinates 
$l$ and $k$ \cite{Ok,Go09}. 
Explicit expressions are for $C_{0,4}$,
\begin{subequations}\label{FN-FK}
\begin{align}
& L_s\,=\,2\cosh(l/2)\,,\\
%\end{align}
%and for $C_e\simeq C_{1,1}$,
%\begin{align}
%& L_t\big((L_s)^2-4\big)^{\frac{1}{2}}\,=\,2\cosh(k/2)
%\sqrt{(L_s^e)^2+L_0^e-2}\\
%& L_u\big((L_s)^2-4\big)^{\frac{1}{2}}\,=\,
%2\cosh((l_e+k_e)/2)
%\sqrt{(L_s^e)^2+L_0^e-2}\,,
%\end{align}
%while 
%\begin{align}
& L_t\big((L_s)^2-4\big)\,=\,
2(L_2L_3+L_1L_4)+L_s(L_1L_3+L_2L_4) \label{cl-'t Hooft}\\
& \hspace{3cm}
+2\cosh(k)
\sqrt{c_{12}(L_s)c_{34}(L_s)}\,,
\notag\\
& L_u\big((L_s)^2-4\big)\,=\,
L_s(L_2L_3+L_1L_4)+2(L_1L_3+L_2L_4) \label{cl-dyonic}\\
& \hspace{3cm}
+2\cosh((2k-l)/2)
\sqrt{c_{12}(L_s)c_{34}(L_s)}\,,
\notag
\end{align}
\end{subequations}
where $L_i=2\cosh\frac{l_i}{2}$, and $c_{ij}(L_s)$ is defined as
\begin{align}\label{cijdef-app}
c_{ij}(L_s) & \,=\,L_s^2+L_i^2+L_j^2+L_sL_iL_j-4\ . 
%\\ \nonumber & =
%2 \cosh \fr{l_s+l_i+l_j}{4} 2 \cosh \fr{l_s+l_i-l_j}{4}
%2 \cosh \fr{l_s-l_i+l_j}{4} 2 \cosh \fr{l_s-l_i-l_j}{4}.
\end{align}
These expressions
ensure that the algebraic relations $P_e(L_s,L_t,L_u)=0$
are satisfied.
By complexifying $(l,k)$ one gets (local) coordinates for 
$\CM_{\rm flat}^{\BC}(C)$ \cite{NRS}.

\subsection{Poisson structure}

\begin{figure}[t]
\epsfxsize8cm
\centerline{\epsfbox{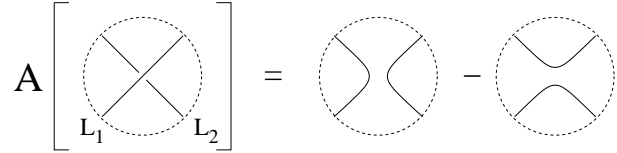}}
\caption{The anti-symmetric smoothing operation}\label{skeinfigas}\vspace{.3cm}
\end{figure}

There is also a natural Poisson bracket on $\CA(C)$ \cite{Go86}, 
defined such that 
\begin{equation}
\{\,L_{\ga_1}\,,\, L_{\ga_2}\,\}\,=\,L_{A(\ga_1,\ga_2)}\,,
\end{equation}
where $A(\ga_1,\ga_2)$ is the loop obtained from 
$\ga_1$, $\ga_2$ by means of the anti-symmetric smoothing operation,
defined as above, but replacing the rule depicted in 
Figure \ref{skeinfig} by the one depicted in Figure 
\ref{skeinfigas}. This Poisson structure coincides with the 
Poisson structure coming
from the natural symplectic structure on $\CM_{\rm flat}(C)$ which
was introduced by Atiyah and Bott.

The resulting expression for the Poisson bracket $\{\,L_s\,,L_t\,\}$
can be written elegantly in the form
\begin{equation}\label{loopPB}
\{\,L_s\,,L_t\,\}\,=\,\frac{\pa}{\pa L_u} P(L_s,L_t,L_u)\,.
\end{equation}
It is remarkable that the same polynomial appears both in 
\rf{algrel} and in \rf{loopPB}, which indicates that the symplectic
structure on  $\CM_{\rm flat}$ is compatible with its 
structure as algebraic variety.

The Fenchel-Nielsen coordinates are known to be
Darboux-coordinates for $\CM_{\rm flat}(C)$, 
having the
Poisson bracket 
\begin{equation}
\{\,l\,,\,k\,\}\,=\,2%\de_{e,e'}
\,.
\end{equation}
The Poisson structure is also rather simple in terms of
the Fock-Goncharov coordinates,
\begin{equation}\label{poisson}
\{\CX_e^{\tau}, \CX_{e'}^{\tau}\}
= n_{e,e'}\, \CX_{e'}^{\tau}\,{\CX_e^{\tau}},
\end{equation}
where $n_{e,e'}$ is the number of faces $e$
and $e'$ have in common, counted with a sign.

% \subsection{Real slice}

% \newpage
\paragraph{\large References to articles in this volume}
\renewcommand{\refname}{

\paragraph{Other references}
\renewcommand{\refname}{

\end{document}
\vskip-36pt}
\begin{thebibliography}{}
% \setcounter{enumiv}{\value{genref}}

\bibitem[V:0]{Int} J. Teschner, {\it Exact results on Supersymmetric Field Theories}.
\bibitem[V:1]{G13} D. Gaiotto, {\it Families of $N=2$ field theories}. 
\bibitem[V:2]{Ne} A. Neitzke, {\it Hitchin systems in $\CN=2$ field theory}.
\bibitem[V:3]{Ta} Y. Tachikawa, {\it A review on instanton counting and
W-algebras}.
\bibitem[V:4]{M} K. Maruyoshi, {\it $\beta$-deformed matrix models and
the 2d/4d correspondence.}
\bibitem[V:5]{PV} V.~Pestun, {\it Localization for $\CN=2$ Supersymmetric
Gauge Theories in Four Dimensions.}
\bibitem[V:6]{O} T. Okuda, {\it Line operators in supersymmetric gauge
theories and the 2d-4d relation}.
\bibitem[V:7]{G} S. Gukov, {\it Surface Operators}.
\bibitem[V:8]{RR} L. Rastelli, S. Razamat, 
{\it Index of theories of class $\CS$: a review}.
\bibitem[V:9]{H} K. Hosomichi, {\it A review on SUSY gauge theories on $S^3$}.
\bibitem[V:10]{D} T. Dimofte, {\it 3d Superconformal Theories from 
Three-Manifolds}.
%\bibitem[V:11]{T} J. Teschner, {\it Supersymmetric gauge theories,
%quantization of $\CM_{\rm flat}$, and Liouville theory.}
\bibitem[V:12]{A} M. Aganagic, {\it Topological strings and 2d/4d 
correspondence}
\bibitem[V:13]{KW} D. Krefl, J. Walcher, {\it B-Model Approaches to 
Instanton Counting}. 
\end{thebibliography}

\vskip-36pt}

\begin{thebibliography}{99}

\newcommand{\CMP}[3]{{ Commun. Math. Phys. }{\bf #1} (#2) #3}
\newcommand{\LMP}[3]{{ Lett. Math. Phys. }{\bf #1} (#2) #3}
\newcommand{\IMP}[3]{{ Int. J. Mod. Phys. }{\bf A#1} (#2) #3}
\newcommand{\NP}[3]{{ Nucl. Phys. }{\bf B#1} (#2) #3}
\newcommand{\PL}[3]{{ Phys. Lett. }{\bf B#1} (#2) #3}
\newcommand{\MPL}[3]{{ Mod. Phys. Lett. }{\bf A#1} (#2) #3}
\newcommand{\PRL}[3]{{ Phys. Rev. Lett. }{\bf #1} (#2) #3}
\newcommand{\AP}[3]{{ Ann. Phys. (N.Y.) }{\bf #1} (#2) #3}
\newcommand{\LMJ}[3]{{ Leningrad Math. J. }{\bf #1} (#2) #3}
\newcommand{\FAA}[3]{{ Funct. Anal. Appl. }{\bf #1} (#2) #3}
\newcommand{\TMP}[3]{{ Theor. Math. Phys. }{\bf #1} (#2) #3}
\newcommand{\PTPS}[3]{{ Progr. Theor. Phys. Suppl. }{\bf #1} (#2) #3}
\newcommand{\LMN}[3]{{ Lecture Notes in Mathematics }{\bf #1} (#2) #2}
\small  \setlength{\itemsep}{-3pt}


\bibitem[AST]{AST} O. Aharony, N. Seiberg, Y. Tachikawa,
{\it Reading between the lines of four-dimensional gauge theories},
JHEP {\bf 1308} (2013) 115.
% [arXiv:1305.0318 [hep-th]]

\bibitem[AFLT]{AFLT}
V.A. Alba, V.A. Fateev, A.V. Litvinov, G.M. Tarnopolsky, 
{\it On combinatorial expansion of the conformal blocks 
arising from AGT conjecture}. 
Lett. Math. Phys. {\bf 98} (2011) 33--64.
%[arXiv:1012.1312 [hep-th]]

%\bibitem[AGMV]{AGMV}
%L. Alvarez Gaum\'e, C. Gomez, G. Moore, C. Vafa,
%{\it Strings in the operator formalism} 
%Nucl. Phys. B {\bf 303} (1988) 455--521



\bibitem[AGT]{AGT}
L.~F.~Alday, D.~Gaiotto, and Y.~Tachikawa,
{\em Liouville Correlation Functions from Four-dimensional Gauge Theories},
Lett. Math. Phys. {\bf 91} (2010) 167--197.
%[arXiv:0906.3219 [hep-th]].

\bibitem[AGGTV]{AGGTV}
L. F. Alday, D. Gaiotto, S. Gukov, Y. Tachikawa, H. Verlinde,
{\em Loop and surface operators in $\mathcal{N}=2$ gauge theory and
Liouville modular geometry}, J. High Energy Phys. {\bf 1001} (2010) 113.
%[arXiv:0909.0945 [hep-th]].

\bibitem[BPZ]{BPZ} 	A.A. Belavin, A.M. Polyakov, 
A.B. Zamolodchikov,
{\it Infinite Conformal Symmetry in Two-Dimensional Quantum Field Theory},
Nucl.Phys. {\bf B241} (1984) 333-380.


\bibitem[BF]{BF} E. Frenkel, D. Ben-Zvi,
{\it Vertex algebras and algebraic curves.} Second edition. 
Mathematical Surveys and Monographs, 88. 
American Mathematical Society, Providence, RI, 2004.


\bibitem[BJSV]{BJSV}
M. Bershadsky, A. Johansen, V. Sadov, and C. Vafa,
{\it Topological Reduction Of 4D SYM To 2D Sigma Models,}
Nucl. Phys. {\bf B448} (1995) 166-186. 
%[arxiv:hep-th/9501096].


\bibitem[BK]{BK} B. Bakalov, A. Kirillov, Jr., 
{ \it On the Lego-Teichm\"uller game.}  
Transform. Groups  {\bf 5}  (2000),  no. 3, 207--244.

%\bibitem[BMW]{BMW} K.
%Burns, H. Masur, A. Wilkinson, 
%{\it The Weil-Petersson geodesic flow is ergodic.}  
%Ann. of Math. (2)  {\bf 175}  (2012) 835--908





\bibitem[CF1]{CF} L.O. Chekhov, V. Fock: {\it A quantum Teich\-m\"uller space},
  Theor. Math. Phys. {\bf 120} (1999) 1245-1259. 
%[arxiv:math/9908165]

\bibitem[CF2]{CF2} L.O. Chekhov, V. Fock: {\it  Quantum modular 
  transformations, the pentagon relation, and geodesics}, 
    Proc. Steklov Inst. Math.  {\bf 226}  (1999)  149-163.

\bibitem[CMS]{CMS} L. Cantini, P. Menotti and D. Seminara, {\it 
Proof of Polyakov conjecture for general
elliptic singularities,} 
Phys Lett. {\bf B517} (2001), 203-209.

\bibitem[CN]{CN} C. Cordova, A. Neitzke, 
{\it Line Defects, Tropicalization, and 
Multi-Centered Quiver Quantum Mechanics.}
JHEP {\bf 09} (2014) 099.  
%[arXiv:1308.6829 [hep-th]]

\bibitem[DGV]{DGV}
T. Dimofte, D. Gaiotto, R. van der Veen,
{\it RG Domain Walls and Hybrid Triangulations},
arXiv:1304.6721 [hep-th].

\bibitem[DMO]{DMO}  
N. Drukker, D.R. Morrison, T. Okuda,
{\it Loop operators and S-duality from curves on Riemann surfaces},
JHEP {\bf 0909} (2009) 031. 
%[arXiv:0907.2593 [hep-th]].

%\bibitem[DS]{DS} J. F. van Diejen and V. P. Spiridonov, {\em Unit circle
%elliptic beta integrals}, Ramanujan J. {\bf 10} (2005) 187--204.


%\bibitem[DHKM]{DHKM}
%N. Dorey, T.J. Hollowood, V.V. Khoze, M.P. Mattis, 
%{\it The Calculus of Many Instantons}. 
%Phys. Reports {\bf 371} (2002) 231-459


\bibitem[DO]{DO}  
H. Dorn, H.-J. Otto 
{\it Two and three-point functions in Liouville theory},
Nucl.\ Phys.\ B {\bf 429} (1994) 375-388.

\bibitem[DGOT]{DGOT} N. Drukker, J. Gomis, T. Okuda, J. Teschner,
{\em Gauge Theory Loop Operators and Liouville Theory},
J. High Energy Phys. {\bf 1002} (2010) 057.
%[arXiv:0909.1105 [hep-th]].

%\bibitem[F2]{F2} L. D.~Faddeev: {\em Discrete Heisenberg--Weyl group 
%and modular group}, \LMP{34}{1995}{249--254}.
 
%\bibitem[FK2]{FK2}
%L.D.~Faddeev and R. M.~Kashaev: {\em Quantum dilogarithm}, 
%Mod.\ Phys.\ Lett.\ {\bf A9} (1994) 427--434.

%\bibitem[FFMP]{FFMP}
%R. Flume, F. Fucito, J. F. Morales, R. Poghossian 
%{\it Matone's Relation in the Presence of Gravitational Couplings}. 
%JHEP 0404 (2004) 008

\bibitem[F97]{F97} V. Fock: {\it Dual Teichm\" uller spaces}
arXiv:dg-ga/9702018.

\bibitem[FG1]{FG}
V.V. Fock, A. Goncharov,
{\it Moduli spaces of local systems and higher Teichmüller theory.} 
Publ. Math. Inst. Hautes \'Etudes Sci. {\bf 103} (2006) 1--211.


\bibitem[FG2]{FG2}  Fock, V. V.; Goncharov, A. B. 
{\it The quantum dilogarithm and representations of 
quantum cluster varieties.} 
Invent. Math. {\bf 175} (2009) 223--286.

%\bibitem[FGT]{FGT} E. Frenkel, S. Gukov, J. Teschner,
%in preparation.


%\bibitem[FS]{FS} D. Friedan, S. Shenker,
%{\it The Analytic Geometry of Two-Dimensional Conformal Field Theory},
%Nucl. Phys. {\bf B281} (1987) 509-545


%\bibitem[FMPT]{FMPT}
%F.Fucito, J.F.Morales, R.Poghossian, A.Tanzini, 
%{\it N=1 Superpotentials from Multi-Instanton Calculus}. 
%JHEP 0601 (2006) 031


\bibitem[G09]{G09} D. Gaiotto,  {\it N=2 dualities},
JHEP {\bf 1208} (2012) 034.
%[arXiv:0904.2715 [hep-th]].

\bibitem[GMN1]{GMN1} D. Gaiotto, G. Moore, A. Neitzke,
{\it Four-dimensional wall-crossing via three-dimensional field theory.} 
Comm. Math. Phys. {\bf 299} (2010) 163--224.
%[arXiv:0807.4723 [hep-th]].

\bibitem[GMN2]{GMN2} D. Gaiotto, G. Moore, A. Neitzke,
{\it Wall-crossing, Hitchin systems, and the WKB approximation.}  
Adv. Math.  {\bf 234}  (2013) 239--403.
%[arXiv:0907.3987 [hep-th]].

\bibitem[GMN3]{GMN3} D. Gaiotto, G. Moore, A. Neitzke,
{\it Framed BPS States},
Adv.Theor.Math.Phys. {\bf 17} (2013) 241-397.  
%[arXiv:1006.0146 [hep-th]].


\bibitem[Go86]{Go86} W. Goldman, {\it 
Invariant functions on Lie groups and Hamiltonian flows 
of surface group representations.}  Invent. Math.  {\bf 85}  (1986) 263--302.

\bibitem[Go88]{Go}  W. Goldman,  
{\it Topological components of spaces of representations.}
Invent. Math. {\bf 93} (1988) 557--607.


\bibitem[Go09]{Go09} W. Goldman, 
{\it Trace Coordinates on Fricke spaces of some simple hyperbolic surfaces},
Handbook of Teichm\"uller theory. Vol. II, 611-684, 
IRMA Lect. Math. Theor. Phys., 13, Eur. Math. Soc., Z\"urich, 2009.


\bibitem[GOP]{GOP} J. Gomis, T. Okuda, V. Pestun, 
{\it Exact Results for 't Hooft Loops in Gauge Theories on $S^4$},
JHEP {\bf 1205} (2012) 141.
%[arXiv:1105.2568 [hep-th]]

\bibitem[GT]{GT} D. Gaiotto, J. Teschner,
{\it Irregular singularities in Liouville theory and 
Argyres-Douglas type gauge theories}, 
JHEP {\bf 1212} (2012) 050. 
%[arXiv:1203.1052 [hep-th]].

%\bibitem[Gu]{Gu} S. Gukov, 
%{\it Quantization via mirror symmetry.}  
%Jpn. J. Math.  {\bf 6}  (2011) 65--119.

%\bibitem[GV]{GV} S. Gukov, E. Witten,
%{\it Branes and quantization}.  
%Adv. Theor. Math. Phys.  {\bf 13}  (2009) 1445--1518.


%\bibitem[HJS]{HJS} L. Hadasz, Z. Jaskolski and P. Suchanek,
%\textit{Modular bootstrap in Liouville field theory}, Phys. Lett.
%\textbf{B685} (2010), 79--85.




%\bibitem[HLP]{HLP} K.~Hosomichi, S.~Lee, and J. Park,
%{\em AGT on the S-duality wall},
%J. High Energy Phys. {\bf 2010}, no. 12, 079.


\bibitem[Hi]{Hi} N. Hitchin, \textit{The self-duality equations on a 
Riemann surface},
Proc. London Math. Soc. \textbf{55} (3) (1987), 59--126.

\bibitem[HH]{HH} N. Hama, K. Hosomichi,
{\it Seiberg-Witten Theories on Ellipsoids}, 
JHEP {\bf 1209} (2012) 033, Addendum-ibid. 1210 (2012) 051.
%[arXiv:1206.6359 [hep-th]].


%\bibitem[HKS1]{HKS1} L. Hollands, C.A. Keller, J. Song, 
%{\it From SO/Sp instantons to W-algebra blocks}. 
%JHEP 1103:053,2011 


\bibitem[HKS]{HKS2}
L. Hollands, C.A. Keller, J. Song
{\it Towards a 4d/2d correspondence for Sicilian quivers},
JHEP {\bf 1110} (2011) 100. 
%[arXiv:1107.0973 [hep-th]].


\bibitem[HV]{HV} V. Hinich, A. Vaintrob, 
{\it Augmented Teichm\"uller spaces and orbifolds.}  
Selecta Math. (N.S.)  {\bf 16}  (2010) 533--629.


\bibitem[IOT]{IOT}
Y. Ito, T. Okuda, M. Taki,
{\it Line operators on $S^1\times R^3$ and quantization of 
the Hitchin moduli space},
JHEP {\bf 1204} (2012) 010. 
%[arXiv:1111.4221 [hep-th]].


\bibitem[Ka1]{Ka1} R.M. Kashaev: 
{\it Quantization of Teich\-m\"uller spaces and 
  the quantum dilogarithm,} 
\LMP{43}{1998}{105-115}. 
%[arxiv:q-alg/9705021].   

%\bibitem[Ka2]{Ka2} R.M. Kashaev:
% {\it Liouville central charge in quantum 
%  Teichmuller theory}, Proc. of the Steklov Inst. of Math. 
%   {\bf 226} (1999) 63-71, hep-th/9811203

%\bibitem[Ka3]{Ka3} R.M. Kashaev: {\it On the spectrum of Dehn twists in 
%  quantum Teich\-m\"uller theory,} Physics and combinatorics, 2000 (Nagoya),  
% 63--81, World Sci. Publishing, River Edge, NJ, 2001.

\bibitem[Ka4]{Ka4} R.M. Kashaev, {\it The quantum dilogarithm and Dehn 
  twists in quantum Teich\-m\"uller theory.}  Integrable structures of 
  exactly solvable two-dimensional models of quantum field theory 
  (Kiev, 2000),  211--221, NATO Sci. Ser. II Math. Phys. Chem., 35,
  Kluwer Acad. Publ., Dordrecht, 2001.



%\bibitem[Ka98]{Ka97} R. M. Kashaev,
%{\it Quantization of Teichm\"uller spaces and the quantum dilogarithm},
%Lett. Math. Phys.  {\bf 43}  (1998),  no. 2, 105--115.

%\bibitem[K2]{K2}  R. Kashaev, 
%{\em The quantum dilogarithm and Dehn twists in quantum Teichmüller theory},
%``Integrable structures of exactly solvable two-dimensional models 
%of quantum field theory'', (Kiev, 2000) 211--221, 
%NATO Sci. Ser. II Math. Phys. Chem., 35, 
%Kluwer Acad. Publ., Dordrecht, 2001.


%\bibitem[KLV]{KLV} R. Kashaev, F. Luo, G.
%Vartanov, {\textit A TQFT of Turaev-Viro type on shaped
%triangulations}, arXiv:1210.8393.



%\bibitem[Kaw]{Ka} S. Kawai,
%{\it The symplectic nature of the space of projective connections
%on Riemann surfaces.}
%Math. Ann. {\bf 305} (1996) 161-182

%\bibitem[LMN]{LMN}
%A.S. Losev, A.V. Marshakov, N.A. Nekrasov,
%{\it Small instantons, little strings and free fermions}.  
%From fields to strings: circumnavigating theoretical physics. 
%Vol. 1,  581--621, World Sci. Publ., Singapore, 2005.

\bibitem[LNS]{LNS}
A.S. Losev, N.A. Nekrasov, S. Shatashvili, 
{\it Testing Seiberg-Witten solution.}  
Strings, branes and dualities (Carg\`ese, 1997),  359--372, 
NATO Adv. Sci. Inst. Ser. C Math. Phys. Sci., 520, 
Kluwer Acad. Publ., Dordrecht, 1999.
%[arxiv:hep-th/9801061]

\bibitem[Ma]{Ma} A. Marden, 
{\it Geometric complex coordinates for Teichm\"uller space}. Mathematical
aspects of string theory, 341--354, Adv. Ser. Math. Phys. {\bf 1}, World Sci.
(1987).

\bibitem[MNS1]{MNS1} G. Moore, N.A. Nekrasov, S. Shatashvili, 
{\it Integrating over Higgs branches.}  
Comm. Math. Phys.  {\bf 209}  (2000)  97--121.
%[arxiv:hep-th/9712241]. 

\bibitem[MNS2]{MNS2} G. Moore, N.A. Nekrasov, S. Shatashvili, 
{\it D-particle bound states and generalized instantons.}  
Comm. Math. Phys.  {\bf 209}  (2000) 77--95.
%[arxiv:hep-th/9803265].

\bibitem[MS]{MS} G. Moore, N. Seiberg,
{\em Classical and quantum conformal field theory},
Comm. Math. Phys. {\bf 123} (1989) 177-254.



\bibitem[N]{N} N.~A.~Nekrasov,
{\em Seiberg-Witten prepotential from instanton counting},
Adv. Theor. Math. Phys. {\bf 7} (2003) 831--864.
%[arxiv:hep-th/0206161]

%\bibitem[NO]{NO} N. Nekrasov, A. Okounkov,
%{\it Seiberg-Witten theory and random partitions}.  
%The unity of mathematics,  525--596, Progr. Math., 244, 
%Birkhäuser Boston, Boston, MA, 2006.

\bibitem[NRS]{NRS} N. Nekrasov, A. Rosly, S. Shatashvili,
{\em Darboux coordinates, Yang-Yang functional, and gauge theory},
Nucl.\ Phys.\ Proc.\ Suppl.\  {\bf 216} (2011) 69--93.
%[arXiv:1103.3919 [hep-th]]. 

%\bibitem[NS98]{NS98}
%N. Nekrasov, A. Schwarz, 
%{\it Instantons on noncommutative $\BR^4$, and $(2,0)$ 
%superconformal six dimensional theory}, 
%Comm. Math. Phys. {\bf 198} (1998) 689--703


\bibitem[NS04]{NS04} N. 
Nekrasov, S. Shadchin,
{\it ABCD of instantons}.  
Comm. Math. Phys.  {\bf 252}  (2004) 359--391.
%[arxiv:hep-th/0404225].

\bibitem[NS]{NS} N.A. Nekrasov, S.L. Shatashvili, 
{\it Quantization of integrable systems and four dimensional gauge theories.} 
XVIth International Congress on Mathematical Physics, 265--289, 
World Sci. Publ., Hackensack, NJ, 2010.
%[arXiv:0908.4052 [hep-th]].

\bibitem[NT]{NT} I. Nidaeiev, J. Teschner, 
{\it On  the relation between the modular double of $\CU_{q}(\fsl(2,\BR))$ 
and the quantum Teichm\"uller theory},
Preprint arXiv:1302.3454

\bibitem[NW]{NW} N. Nekrasov, E. Witten,
{\it The Omega Deformation, Branes, Integrability, and Liouville Theory},
JHEP {\bf 1009} (2010) 092. 
%[arXiv:1002.0888 [hep-th]].

\bibitem[Ok]{Ok} T. Okai, 
{\it Effects of change of pants decomposition on their
Fenchel-Nielsen coordinates}, 
Kobe J. Math. {\bf 10} (1993) 215-223.

\bibitem[Pe]{Pe} V.~Pestun, {\em Localization of gauge theory on
a four-sphere and supersymmetric Wilson loops}, 
Comm. Math. Phys. {\bf 313} (2012) 71--129.
%[arXiv:0712.2824 [hep-th]].

%\bibitem[PT1]{PT1}
%B.~Ponsot and J.~Teschner,
%{\em Liouville bootstrap via harmonic analysis on a noncompact quantum group}, {\tt arXiv:hep-th/9911110}.

%\bibitem[PT2]{PT2}
%B.~Ponsot and J.~Teschner,
%{\em Clebsch-Gordan and Racah-Wigner coefficients for a continuous series of representations of $U_q(sl(2,\mathbb{R}))$}, Commun.\ Math.\ Phys.\  {\bf 224} (2001) 613--655.

\bibitem[RS]{RS}
J.W. Robbin, D.A. Salamon,
{\it A construction of the Deligne-Mumford orbifold.} 
J. Eur. Math. Soc. (JEMS) {\bf 8} (2006) 611--699.
%[arXiv:math/0407090].


%\bibitem[Ru]{Ru} S. N. M.~Ruijsenaars: {\em First order analytic 
%difference equations and integrable quantum systems},
%J.~Math.\ Phys.\ {\bf  38} (1997) 1069--1146.

%\bibitem[SchV]{SchV} O. Schiffmann, E. Vasserot
%{\it Cherednik algebras, W algebras and the equivariant 
%cohomology of the moduli space of instantons on ${\mathbb A}^2$}. 
%Preprint arXiv:1202.2756 

%\bibitem[S01]{S1}  V. P. Spiridonov,
%{\em On the elliptic beta function}, Uspekhi Mat. Nauk {\bf 56}
%(1) (2001) 181--182 (Russian Math. Surveys {\bf 56} (1) (2001) 185--186).

%\bibitem[S03]{S2} V. P. Spiridonov, {\em Theta hypergeometric
%integrals}, Algebra i Analiz {\bf 15} (6) (2003) 161--215 (St.
%Petersburg Math. J. {\bf 15} (6) (2004), 929--967).

%\bibitem[S08]{S3} V. P. Spiridonov, {\em Essays on the theory of
%elliptic hypergeometric functions}, Uspekhi Mat. Nauk {\bf 63} (3)
%(2008), 3--72 (Russian Math. Surveys {\bf 63} (3) (2008), 405--472).

%\bibitem[SV10]{SV0}
%V.~P.~Spiridonov and G.~S.~Vartanov,
%{\em Superconformal indices for $\mathcal{N} = 1$ theories with multiple duals}, Nucl.\ Phys.\ B {\bf 824} (2010) 192--216.

%\bibitem[SV11]{SV11}
%V.~P.~Spiridonov and G.~S.~Vartanov, {\em Elliptic hypergeometry of supersymmetric dualities II. Orthogonal groups, knots, and vortices}, {\tt arXiv:1107.5788 [hep-th]}.

%\bibitem[Sp]{Sp} M. Spreafico,
%{\em On the Barnes double zeta and Gamma functions},
%Journal of Number Theory {\bf 129} (2009) 2035-2063.

% \bibitem[TZ85]{TZ85} P.G. Zograf and L.A. Takhtajan,
%\textit{Action of the Liouville equation as generating
%function for accessory parameters and the potential of the
%Weil--Petersson metric on Teichm\"uller space}, Funct.
%Anal. Appl. \textbf{85} (1985), 218--220.

\bibitem[Ta13]{Ta13} Y. Tachikawa, 
{\it On the 6d origin of
discrete additional data of 4d gauge theories},
JHEP {\bf 1405} (2014) 020.
%[arXiv:1309.0697 [hep-th]].

\bibitem[TZ87a]{TZ87a} L.A. Takhtajan, P.G. Zograf,
\textit{On the Liouville equation, accessory parameters and
the geometry of Teichm\"uller space for Riemann surfaces of
genus 0}, Math. USSR-Sb.  \textbf{60} (1988) 143--161.

%\bibitem[TZ87b]{TZ87b} P.G. Zograf and L.A. Takhtajan,
%\textit{On the uniformization of Riemann surfaces and on
%the Weil--Petersson metric on the Teichm\"uller and
%Schottky spaces}, Math. USSR-Sb. \textbf{60} (1988), 297--313.

\bibitem[TZ03]{TZ03} L.A. Takhtajan, P.G. Zograf,
\textit{Hyperbolic 2-spheres with conical singularities,
accessory parameters and K\"ahler metrics on $\CM_{0,n}$},
Trans. Amer. Math. Soc.  \textbf{355}(5) (2003) 1857--1867.


%\bibitem[TT03]{TT} L. Takhtajan, L.P. Teo,
%{\it Liouville action and Weil-Petersson metric on deformation spaces, 
%global Kleinian reciprocity and holography.}  
%Comm. Math. Phys.  {\bf 239}  (2003) 183-- 240




\bibitem[T01]{T01}
J.~Teschner, {\em Liouville theory revisited}, Class.\ Quant.\ Grav.\  {\bf 18} (2001) R153--R222.
%[arXiv:hep-th/0104158]

%\bibitem[T03]{T03} 
%J.~Teschner, {\em On the relation between quantum Liouville
%theory and the quantized Teichm\"uller spaces}, Int.\ J.\ Mod.\ Phys.\
%{\bf A19S2} (2004), 459--477; {\em From Liouville theory to the
%quantum geometry of Riemann surfaces}, Cont. Math. {\bf 437} (2007) 231--246.

\bibitem[T03a]{T03a} J. Teschner,
{\it A lecture on the Liouville vertex operators},
Int. J. Mod. Phys. {\bf A19S2} (2004) 436-458.
%[arXiv:hep-th/0303150].

\bibitem[T03b]{T} J. Teschner,
{\it From Liouville theory to the quantum geometry of
Riemann surfaces.}
Prospects in mathematical physics,  231--246, Contemp. Math., 437,
Amer. Math. Soc., Providence, RI, 2007.
%[arXiv:hep-th/0308031].


\bibitem[T05]{T05} J. Teschner,
{\em An analog of a modular functor from quantized Teichm\"uller 
theory}, ``Handbook of Teichm\"uller theory'', (A. Papadopoulos, ed.) 
Vol. I, EMS Publishing House, Z\"urich 2007, 685--760.
%[arXiv:math/0510174]. 


%\bibitem[T08]{T08}  J. Teschner, {\em Nonrational conformal field theory}, 
%``New Trends in Mathematical Physics'' 
%(Selected contributions of the XVth ICMP),
%Vladas Sidoravicius (ed.), Springer Science and Business Media B.V. 2009.

\bibitem[T10]{T10} J. Teschner,
{\it Quantization of the Hitchin moduli spaces, Liouville theory,
and the geometric Langlands correspondence I}.
 Adv. Theor. Math. Phys.  {\bf 15}  (2011) 471--564.
%[arXiv:1005.2846 [hep-th]].

\bibitem[TV12]{TV} J. Teschner, G. S. Vartanov,
{\it 6j symbols for the modular double, quantum 
hyperbolic geometry, and supersymmetric gauge theories}.
Lett. Math. Phys. {\bf 104} (2014) 527-551. 
%[arXiv:1202.4698 [hep-th]].

\bibitem[TV13]{TV2} J. Teschner, G. S. Vartanov,
{\it Supersymmetric gauge theories, quantization of 
moduli spaces of flat connections, and conformal field theory},
arXiv:1302.3778. 
 
%\bibitem[V]{V}
%A.Yu.~Volkov: {\em Noncommutative hypergeometry}, \CMP{258}{2005}{257--273}.

%\bibitem[Wi88]{W88}
%E. Witten, {\it Quantum field theory, Grassmannians, and algebraic curves.}  
%Comm. Math. Phys.  {\bf 113}  (1988) 529--600.


%\bibitem[Wo]{Wo}
%S.L.~Woronowicz: {\em Quantum exponential function}, 
%Rev.\ Math.\ Phys.\ {\bf 12} (2000) 873--920.

\bibitem[ZZ]{ZZ}  A.B.Zamolodchikov, Al.B.Zamolodchikov,
{\it Structure Constants and Conformal Bootstrap in Liouville Field Theory.}
Nucl. Phys. {\bf B477} (1996) 577--605.
%[arxiv:hep-th/9506136].

%\bibitem[Zo]{Zo} P.G. Zograf,
%{\it The Liouville Action on Moduli Spaces, 
% and Uniformization of Degenerating 
% Riemann Sufraces},
%Leningrad Math. J. {\bf 1} (1990) 941 -- 965

\end{thebibliography}
